\documentclass[epsfig, usenatbib, usegraphicx]{mn2e}
\usepackage{epsfig}
\usepackage{natbib}
\title{Non-Gaussianity in the WMAP data using the peak-peak correlation function}
\author[Tojeiro et al.]{R. Tojeiro\thanks{E-mail: rmft@roe.ac.uk}$^1$, P. G.
Castro$^1$, A. F. Heavens$^1$ and S. Gupta$^2$ \\
$^1$Institute for Astronomy, University of Edinburgh, Royal
Observatory, Blackford Hill, Edinburgh, EH9 3HJ, UK\\
$^2$Institute of Cosmology $\&$ Gravitation,  Mercantile House,
University of Portsmouth, Portsmouth, PO1 2EG, UK}

\def\gs{\mathrel{\raise1.16pt\hbox{$>$}\kern-7.0pt %
\lower3.06pt\hbox{{$\scriptstyle \sim$}}}}         %
\def\ls{\mathrel{\raise1.16pt\hbox{$<$}\kern-7.0pt %
\lower3.06pt\hbox{{$\scriptstyle \sim$}}}}         %

\newcommand{\AaA}{A\&A}
\newcommand{\AJ}{ApJ}
\newcommand{\AJS}{ApJS}

\newcommand{\AJL}{ApJL}

\newcommand{\MNRAS}{MNRAS}
\newcommand{\Nat}{Nat}
\newcommand{\NewAst}{New Astronomy}

\newcommand{\PRD}{Phys. Rev. D}

\begin{document}

\maketitle

\begin{abstract}
We present a search for non-Gaussianity in the WMAP first-year data using
the two-point correlation function of maxima and minima in the temperature
map.  We find evidence for non-Gaussianity on large scales, whose origin
appears to be associated with unsubstracted foregrounds, but which is not
entirely clear.  The signal appears to be associated most strongly with
cold spots, and is more pronounced in the Southern galactic hemisphere.
Removal of the region of sky near the galactic plane, or filtering out
large-scale modes removes the signal.  Analysis of individual frequency
maps shows strongest signal in the 41GHz Q band. A study of difference
maps tests the hypothesis that the non-Gaussianity is due to residual
foregrounds and noise, but shows no significant detection.  We suggest
that the detection is due to large-scale residual foregrounds affecting more than one
frequency band, but a primordial contribution from the Cosmic Microwave
Background cannot be excluded.

\end{abstract}

\begin{keywords}
cosmic microwave background -- early universe -- methods: data
analysis -- methods: numerical -- methods: statistical
\end{keywords}

\title{Non-Gaussianity in the WMAP data using the peak-peak correlation function}

\section{Introduction}

The hypothesis of primordial statistical random Gaussianity lies at
the core of our understanding of the Universe. The currently
favoured standard cosmological model is built upon the assumption of
a set of initial conditions laid down during
Inflation \citep{art:Guth81}. In the most simple model of Inflation,
these initial conditions produce nearly Gaussian temperature
fluctuations of the Cosmic Microwave Background (CMB). If the CMB
temperature field is indeed Gaussian distributed, its two-point
angular correlation function (or power spectrum in dual space)
contains all the information needed to characterise its statistical properties fully. In
the Inflation paradigm and the hypothesis of statistical isotropy
(i.e. invariance under rotations of n-point correlation functions),
the angular power spectrum $C_{\ell}$ naturally became the standard
statistical tool from which to extract all the cosmological
information present in the CMB data~\citep{art:JungmanEtAl96}. If, on
the contrary, the microwave background radiation is significantly
non-Gaussian, it would be powerful evidence against Inflation with
far-reaching consequences for our understanding of structure
formation. 

As a consequence, testing the precise statistical nature of the CMB
anisotropies is of extreme importance in order to test the validity
of our present theories of structure formation and of our current
methods of parameter estimation. These statistical tests are
complicated by the inevitable presence of foregrounds, such as the
emission from our own galaxy or extra-galactic point sources and
systematic and instrumental effects which may introduce spurious
non-Gaussianities in the measurements. At small angular scales,
secondary anisotropies such as the Sunyaev-Zel'dovich
effect~\citep{art:SunyaevZeldovich72}, the Ostriker-Vishniac
effect~\citep{art:OstrikerVishniac86,art:Castro03}, the Rees-Sciama
effect~\citep{art:ReesSciama68} and lensing by intervening
structures~\citep{art:BlanchardSchneider87,art:Hu01,art:CoorayKesden03}
as well as cross correlations between
them~\citep{art:CoorayHu00,art:Cooray01a,art:Cooray01b,thesis:Komatsu02,art:VerdeSpergel02}
will also leave their own non-Gaussian signatures; these a
valuable source of cosmological and astrophysical information in
their own right.

Due to the scientific importance of measuring any non-Gaussianity,
but also to the difficulty of detecting it, many non-Gaussianity
studies have been performed at various angular scales since the
first CMB measurements by the Cosmic Background Explorer (COBE)
satellite~\citep{art:SmootEtAl92}. Remarkably COBE provided the first
non-Gaussianity detection by means of the bispectrum
\citep{art:FerreiraEtAl98,art:Heavens98}, a detection that was later claimed to be
due to undocumented systematic
effects~\citep{art:BandayEtAl99,art:Magueijo00,art:MagueijoMedeiros03}. With the
advent of higher sensitivity and higher resolution experiments,
in particular the Wilkinson Microwave Anisotropy Probe
(WMAP; ~\citealt{art:BennettEtAl03}), with its low noise and large sky
coverage, the pursuit of non-Gaussianity has become one of the
main objectives of the CMB field.

As non-Gaussianity can take innumerable forms, it is difficult to test
its presence and there is no optimal general test for detecting and
quantifying it. Different tests can show higher or lower sensitivity
depending on the type of non-Gaussianity being tested (see e.g.
~\citet{art:AghanimEtAl03}). Clearly, it is important to apply as
many different estimators as possible as the eventual understanding
of any deviation from Gaussianity will likely come from a
combination of complementary results.

A multitude of non-Gaussian estimators have thus been applied to the
WMAP first year data such as the bispectrum, Minkowski functionals,
three-point angular correlation function, higher-order moments of
the angular correlation function, wavelets, phase correlations,
tensor modes etc.
\citep{art:ColleyEtAl03,art:KomatsuEtAl03,art:Park03,art:VielvaEtAl03,art:ColesEtAl04,art:CopiEtAl04,art:CruzEtAl04,art:EriksenEtAl03,art:EriksenEtAl04,art:EriksenEtAl04b,art:LandMagueijo04b,art:LandMagueijo04,art:McEwenEtAl04,art:MukherjeeWang04,art:GurzadyanEtAl05,art:JaffeEtAl05,art:LandMagueijo05a,art:LiuZhang05}.
Curiously, the data consistently
show a variety of anomalies. The first of them concerns the low
value of the quadrupole (multipole $\ell=2$), previously observed
in the COBE data and whose origin has been discussed in terms of
super-horizon fluctuations and spatial curvature e.g. \citet{art:BereraEtAl98,art:BereraHeavens00,art:Efs03}. 
Subsequently, studies indicated an unusual low
probability planarity and alignment of the quadrupole and of the
octopole ($\ell=3$), maybe even extending to higher
multipole values ($\ell=5$). In addition, an analysis performed with Spherical
Mexican Hat Wavelets~\citep{art:CruzEtAl04} provided the detection of
a very non-Gaussian structure in the data, associated to a cold spot
of around $10$ degrees, visible in the WMAP temperature maps.
Last but not least, the data appear to exhibit a North-South
asymmetry on large scales. Notoriously, all these claims indicate
not only a deviation from the Gaussianity hypothesis but most
alarming a violation of statistical isotropy, one of the two
fundamental pillars -- together with statistical homogeneity-- of
the Cosmological Principle. A plausible explanation for
these surprising results is likely to be simply the presence of
unaccounted systematics. But one cannot put aside the possibility of
there being of cosmological origin, in which case it would have
far-reaching consequences for our perception of the Universe and
thus for the standard Cosmological model.

The peak-peak correlation function of a Gaussian random field has
been calculated exactly, both in the full-sky and in the flat-sky
approximation, and depends, as it must, only on the power
spectrum~\citep{art:HeavensSheth99,art:HeavensGupta01}. It has been
shown to display a rich structure, in particular at small angular
scales, raising the hope that it has a good sensitivity to a global
non-Gaussian signal. The present study uses an
estimator of the 2-point correlation function of the temperature hot spots
(local maxima) and cold spots (local minima) to look for any
deviation from the Gaussian hypothesis in the WMAP first-year
temperature data. Following the claims of a North-South asymmetry,
we also analyse both hemispheres separately.

This paper is organised as follows. In Section~\ref{sec:method} 
and~\ref{sec:gaussian_maps} we explain in detail our methodology, including the map 
construction pipeline, the peak-peak
correlation function estimator used and the statistical 
method chosen to test robustly the Gaussianity of the data. 
In Section~\ref{sec:results} we apply our method to the 
WMAP first year maps and present our results. 
In order to test various statistical properties of the whole data, we
performed a full-sky study and a North-South comparison analysis of 
a map resulting from combining all the different frequency assemblies
maps provided by WMAP. 
In particular we investigated how this North-South 
comparison changes with cuts both in harmonic
and real space. To further test for the origin of any non-Gaussianity,
we present both a full-sky and a North-South analysis 
performed on single-frequency maps and on frequency-difference maps,
which removes the primordial signal and tests noise and residual foregrounds. 
Finally, in Section~\ref{sec:conclusions} 
we present our conclusions.

%%%%%%%%%%%%% Section II %%%%%%%%%%%%%%%%%%%

\section{Method}
\label{sec:method}

The two-point correlation function of peaks of a Gaussian random field
can be calculated analytically, given only its power spectrum
~\citep{art:HeavensGupta01}. 

The power spectrum has been estimated from the WMAP
data \citep{art:HinshawEtAl03}, and we test the hypothesis that the
field giving rise to this power spectrum is Gaussian. 
We do this by taking a brute force approach and use the best-fit
$\Lambda$CDM theoretical power spectrum to the WMAP data as the starting point to
generate a large number of Gaussian maps at all WMAP frequencies, to
which we then apply sky cuts, window functions and noise so as to
mimic the real data.

In this way, we estimate the hot spot and cold spot correlation
functions, and their covariance, for the Gaussian maps created with
the same algorithm as the real data.  We then use these to test
the Gaussian hypothesis.

\subsection{The maps}\label{sec:maps}

\subsubsection{The WMAP map}

The WMAP satellite probed the CMB at five different frequencies with
two radiometers, producing ten differencing assemblies (DAs): four
on the W-band (94GHz), two on the V-band (61GHz), two on the Q-band
(41GHz), one on the Ka-band (33GHz) and one on the K-band (23GHz).
Each of these assemblies, after calibration and removal of the
monopole and dipole, is available for download from the WMAP web-site\footnote{http://lambda.gsfc.nasa.gov/product/map/m\_products.cfm}. 
All the maps are
provided in the Hierarchical Equal Area isoLatitude Pixelisation
(HEALPix) scheme\footnote{http://www.eso.org/science/healpix/},
which has proved to have several advantages over other methods of
pixelising the surface of a sphere, in particular the fact that the
pixel area is kept constant throughout the surface of the sphere.
However, the pixel shapes can vary largely between the equatorial
and polar regions and distance between pixel centres is not kept
constant. The HEALPix scheme divides the sphere surface into 12
faces of 4 sides each, giving a minimum resolution of 12 pixels.
Each side is divided in $N_{side}$ pixels, giving a total number of
pixels in a map of $12N_{side}^2$. The WMAP maps were provided at a
resolution of $N_{side} = 512$ giving a total of 3,145,728 pixels
separated on average by $\theta_{pix} = 0.115$ degrees $=6.87$
arc minutes.

Each DA map pixel $p$ contains the temperature field (in mK) and a field
containing the number of observations, $N_{obs}(p)$, which allows the
noise per pixel to be estimated using
\begin{equation}
\sigma(p) = \frac{\sigma_0}{\sqrt{N_{obs(p)}}}
\label{eq:noise}
\end{equation}
where $\sigma_0$ is the noise dispersion per map and which has been
published for each of the different assemblies \citep{art:HinshawEtAl03}. Also available
is a foreground-cleaned map of each of the DAs (for details concerning
the removal of the foregrounds see \citet{art:BennettEtAl03}), from which a
Galactic foreground template has been removed, consisting of
synchrotron, free-free and dust emission.

The WMAP map used in this work is a linear combination of the eight
foreground-cleaned assemblies in the Q, V and W bands (the QVW map). We follow the
construction method suggested by 
\citet{art:KomatsuEtAl03} and produce a temperature map given by
\begin{equation}
T_{QVW}(p) = \frac{\sum^8_{j=1}T_j(p)w_j(p)}{\sum_{j=1}^8w_j(p)}
\label{eq:QVW_map}
\end{equation}
with the weights being given by
\begin{equation}
w_j(p) = \frac{1}{\sigma_j^2(p)}
\label{eq:QVW_weights}
\end{equation}
The index $j$ corresponds to the different DAs: $j=1,2$ corresponds to
the V band, $j=3,4$ to the Q band and $j=5$ to $8$ to the W band.

This map proved to be well suited for this work. Compared to other
foreground-cleaned maps such as the Internal Linear Combination
(ILC) map published by the WMAP team and the Tegmark Clean Map
\citep{art:TegmarkEtAl03} (TCM), the QVW map keeps all the small scale
structure by not smoothing the DAs to a common resolution before
construction (also keeping the noise per pixel nicely uncorrelated)
and shows less noise power at high multipoles. And since all the
weights are known in the QVW map, it also becomes trivial to
construct Gaussian simulations of this map.

\subsubsection{The Gaussian maps}

\begin{figure}
\vspace{0.6cm}
\begin{center}
\includegraphics[width=2.0in]{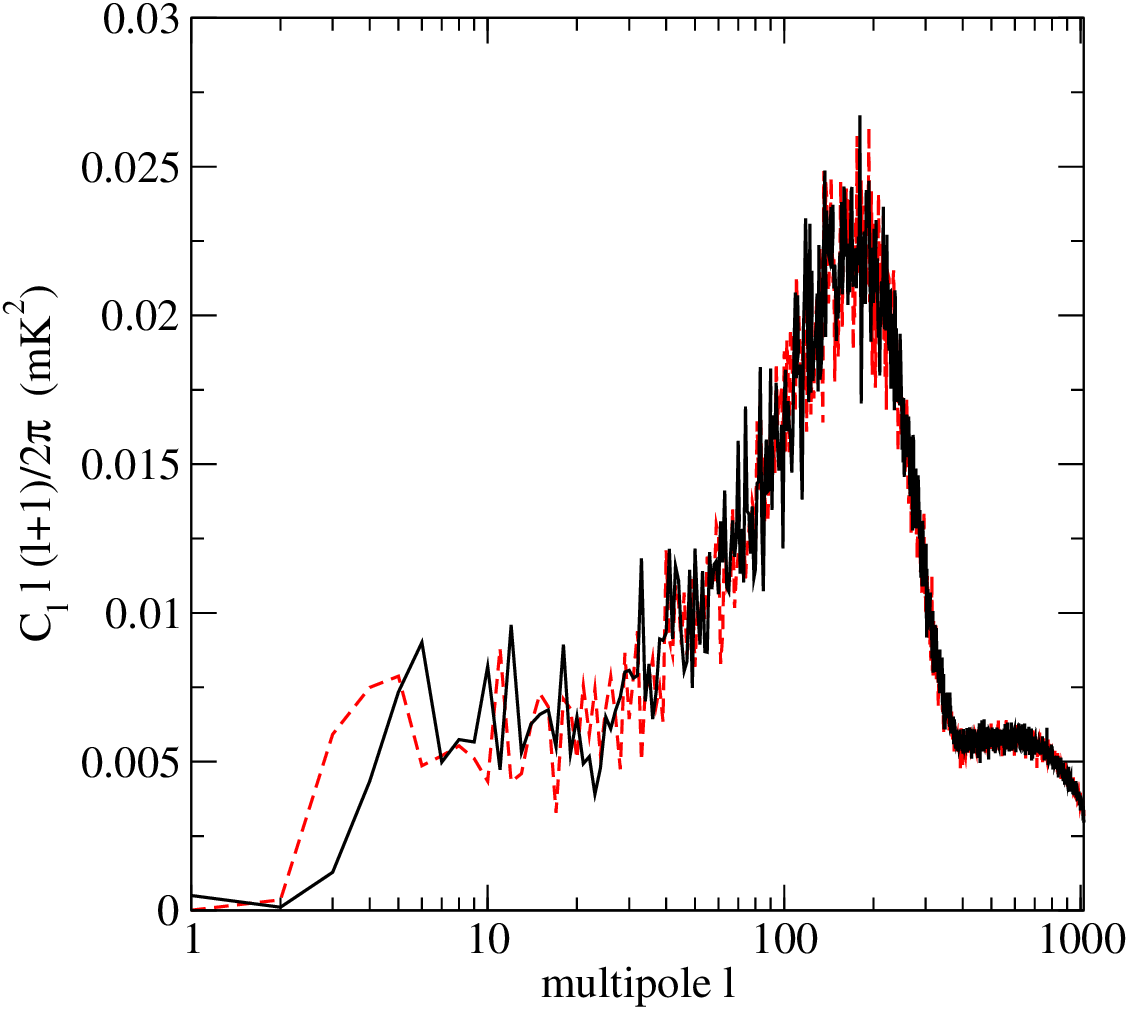}
\caption{\small The power spectrum of our working maps. WMAP data in
the solid line (black in the online version) and one of our Gaussian maps in the dashed
line (red in the online version).}
\label{fig:power_spectra_comp}
\end{center}
\end{figure}

To construct Gaussian simulations of the CMB, we follow the method
suggested by \citet{art:KomatsuEtAl03} and proceed in the following way:
\begin{itemize}
\item We generate one sky realisation from the best fit $\Lambda$CDM model power spectrum, published
in the WMAP web-site.
\item We copy this map 8 times, one for each assembly, and convolve each of the copies with the appropriate
beam transfer function, again published in the WMAP web-site.
\item We add uncorrelated noise to each of the maps according to equation
(\ref{eq:noise}) (a more accurate noise model is used for difference maps). 
\item We combine the 8 resulting maps using equations
(\ref{eq:QVW_map}) and (\ref{eq:QVW_weights}).
\end{itemize}
We repeat this procedure to create many Gaussian simulations of the
CMB, each being a random Gaussian realisation of the same initial
power spectrum. We used different numbers of Gaussian maps in different types of 
analysis, and we quote each number within the appropriate
section. The maps are time-consuming to produce but in each case we check
convergence of $\chi^2$ (see Figures \ref{fig:convergence_fs},
\ref{fig:convergence} and \ref{fig:convergence_QW} for examples). A comparison of the power spectrum of the real and
a simulated map can be seen in Figure \ref{fig:power_spectra_comp}.

Although at small angular scales the
noise properties are white, fully understood and easily
modelled (see discussion in section 2.1.1), at large angular scales,
individual Q, V and W assemblies present noise
characteristics which are non-white. Fortunately these are entirely dominated by the
signal and one does not need to worry about them (see Fig. 1 of
~\citet{art:HinshawEtAl03}). 
The WMAP team have produced a set of 110 noise maps which
include white noise (dominating at small scales), $1/f$ noise
(dominating at large scales) and inter-channel correlations for each
of the radiometers, and ideally one would like to incorporate all
known effects into the analysis. However, being limited by the
relatively small number of full noise simulations and due to the
high signal to noise ratio at the scales where the noise properties
deviate from white, we choose to include only white noise in our
Gaussian QVW and single frequency maps.

\subsubsection{A more extensive set of masks}

In addition to the masks published in the WMAP web-site, a new set of
masks was constructed, based on the WMAP team's kp0 mask which masks
the galactic plane plus all known point sources in the sky.
The initial motivation came from the need to work with smoothed
maps. The QVW maps (real and Gaussian) show a large number of peaks
at full resolution which creates a problem concerning the computational
time taken to calculate the peak-peak correlation function. 
So a trade-off between resolution and computational
limitations was reached and this consisted of applying an additional
Gaussian beam of FWHM = 12 arc minutes to the raw maps (real and Gaussian), 
as constructed in Section 2.1.1. Because some of the 
unwanted (masked) signal could leak from the
masked regions due to this smoothing, the mask kp0 was also
increased (by effectively smoothing it with the same smoothing
kernel applied to the data) to create a new mask, dubbed mask
30B. Regions in the smoothed mask with values greater than 0.01 were
masked, where the raw mask map has values zero and one, with one meaning
masked. \\

In addition to this extension, the new mask was further modified to
include either only the Northern or the Southern hemisphere and also to
optionally exclude a region close the galactic plane, $|b|<30^\circ$.\\

A summary of the properties of each of the new masks is given in Table \ref{tab:mask}.

\begin{table}
\caption{\small The new and original masks}
\begin{center}
\begin{tabular}{|c|c|c|c|} \hline
Mask name      & $|b|$ cut  & Percentage of removed sky  \\
               & (degrees)  &                        \\ \hline
kp0            &    -       &    23                  \\
30B            &    -       &    28                  \\
30B\_knorth    &    -       &    64                  \\
30B\_ksouth    &    -       &    64                  \\
bcut30\_knorth &    30      &    76                  \\
bcut30\_ksouth &    30      &    77                  \\
\hline
\end{tabular}
\end{center}
\label{tab:mask}
\end{table}

\subsection{Peak-peak correlation function}

The two-point correlation function {\bf $\xi$} 
of a discrete distribution of points can be defined as
\begin{equation}
dP = n(1+\xi(\theta))d\Omega
\end{equation}
where $dP$ is the probability of finding a data point within a
solid angle $d\Omega$, at a distance $\theta$ from another randomly chosen data
point and $n$ is the average density of points in the sample. So $\xi(\theta)$ can
be seen as an {\it excess} of probability of finding a pair at a distance
$\theta$, compared to a random catalogue - it is a measurement of clustering - and
$\xi(\theta) = 0$ indicates a uniform random distribution (no clustering). \\

\par\noindent
There are
several estimators suggested in the literature to estimate $\xi(\theta)$
directly from the data. They all work by comparing the sample of
points to an uniform, random catalogue with the same spatial
distribution as the real data. We used the \citet{art:Hamilton93} estimator, which promises fast convergence:
\begin{equation}
\xi(\theta) = \frac{RR(\theta).DD(\theta)}{DR(\theta)^2} - 1 \label{eq:ham}
\end{equation}
where $RR(\theta)$ and $DD(\theta)$ are the number of random and data
pairs respectively at a distance $\theta$ from each other and $DR(\theta)$ is
the number of cross-pairs separated by a distance $\theta$ (all
weighted by the number of total random, data and cross pairs in the
catalogue). Indeed, we found it to converge faster than the standard
estimator, $\xi(\theta) = \frac{DD(\theta)}{DR(\theta)}-1$. We
use large random catalogues of 400,000 points for full sky
and 200,000 points for single hemispheres, with the same sky cut as
the appropriate WMAP map, and ensure that the estimator has
converged to a stable value. A hot spot (cold spot) is defined for the
purposes of this analysis as the centre of any pixel whose temperature
is higher (lower) than the temperature of all pixels with which it shares a boundary.

The correlation function is estimated in 300 
equally-spaced bins up to a maximum separation of 1800
arc minutes.

%%%%%%%%%%%%% Section III %%%%%%%%%%%%%%%%%%%

\section{Comparison with Gaussian maps}
\label{sec:gaussian_maps}

We test the Gaussianity hypothesis of the WMAP data by comparing
our estimator of the peak-peak correlation function $\xi(\theta)$
applied to the data with its values when applied to 
a set of synthetic Gaussian maps with the same cosmology as the WMAP best-fit.

As explained in the previous section,
we apply an additional Gaussian beam of FWHM of 12 arc minutes to our
real and Gaussian maps, constructed as in section 2.1.1 and 2.1.2, 
to smooth out some of the small scale noise
and make the number of peaks slightly more manageable. We mask all
maps with the mask 30B and we select all maxima with
temperature above a given threshold $\nu$ and all minima below
$-\nu$ ($\nu$ in units of the map rms, $\sigma$). We estimate the
peak-peak correlation function using equation (\ref{eq:ham}).

We choose to quantify the robustness of any non-Gaussianity detection 
by means of the $\chi^2$ statistic, 
which we compute for the WMAP data and for each of the Gaussian maps:
\begin{equation}
\chi^2 = \sum_{i,j} (\xi_{i} - \bar{\xi}^G_i)
C_{ij}^{-1}(\xi_{j} - \bar{\xi}_j^G)
\label{eq:chi_sq}
\end{equation}
where the covariance matrix $C_{ij}$ is estimated from the Gaussian
maps available:
\begin{equation}
C_{ij} = \langle (\xi^G_{i,n} - \bar{\xi}^G_i)(\xi^G_{j,n} - \bar{\xi}^G_j) \rangle
\label{eq:Cij}
\end{equation}
Here $i, j$ are bins at a given angular separation and the $\bar{\xi}_i$ is
the average of our estimator over all Gaussian maps of that particular
bin. Previously to computing equations (\ref{eq:chi_sq}) and
(\ref{eq:Cij}) we rebin all data to 19 bins, of which we discard the
first one\footnote{HEALPix defines neighbouring pixels as
ones which share a pixel face. However, due to the highly variable
pixel shapes in the surface of the sphere, these are not necessarily the
closest pixels to the central one. This occasionally results in HEALpix selecting two very close pixels as being
separate peaks which in turn results in unexpected (but explainable)
features in the first few bins. 
Hence we choose to ignore these bins (which fall into the first one
after rebinning). The effect these extra peaks have at large angular
scales was tested for and found to be negligible.}. Rebinning is
necessary, otherwise $C_{ij}$ is close to singular and numerically
unstable to inversion. 

For Gaussian and uncorrelated errors we recall that the $99\%$
confidence level for a reduced $\chi^2$ distribution with 18 degrees
of freedom comes at $\chi^2 \le 0.38$ and $\chi^2 \ge 1.93$.
We also use the $\chi^2$ distribution estimated directly from all of the Gaussian maps
to give more empirical values for confidence levels - see Section
\ref{sec:summary} for a more detailed discussion on confidence
levels. We present our all results using reduced values of $\chi^2$ with a total of 18 degrees
of freedom.

%%%%%%%%%%%%% Section IV %%%%%%%%%%%%%%%%%%%

\section{Results}
\label{sec:results}

We use the peak-peak correlation function in a number of different ways 
to investigate the properties of the maps (we use H for Hot, C for
Cold, N for North and S for South):
\begin{itemize}
\item The most obvious way is to
conduct a full-sky analysis in the unmasked regions of the maps, which 
we do for hot and cold spots separately - $\xi_{H,i}$ and
$\xi_{C,i}$. 
\item We also compute the peak-peak correlation function
in each of the hemispheres individually, again looking at hot and cold spots
separately in each case - $\xi_{NH,i}$, $\xi_{NC,i}$, 
$\xi_{SH,i}$ and $\xi_{SC,i}$.
\item In addition we look
at the difference of correlation between the two hemispheres at a 
given angular scale and we define $\Delta\xi_{H,i} =
\xi_{SH,i} - \xi_{NH,i}$ (similar for cold spots).
\item Finally, we take the average of the peak-peak correlation
function in the Northern and Southern hemispheres in order to
produce a computationally faster way to estimate the full-sky
function - $\tilde{\xi}_{H,i}$  (similarly for cold spots).
\end{itemize}
We explain each of these in detail in the following sections.

\subsection{All-sky analysis}\label{sec:all_sky}

\begin{figure}
\vspace{0.6cm}
\begin{center}
\includegraphics[width=2.in]{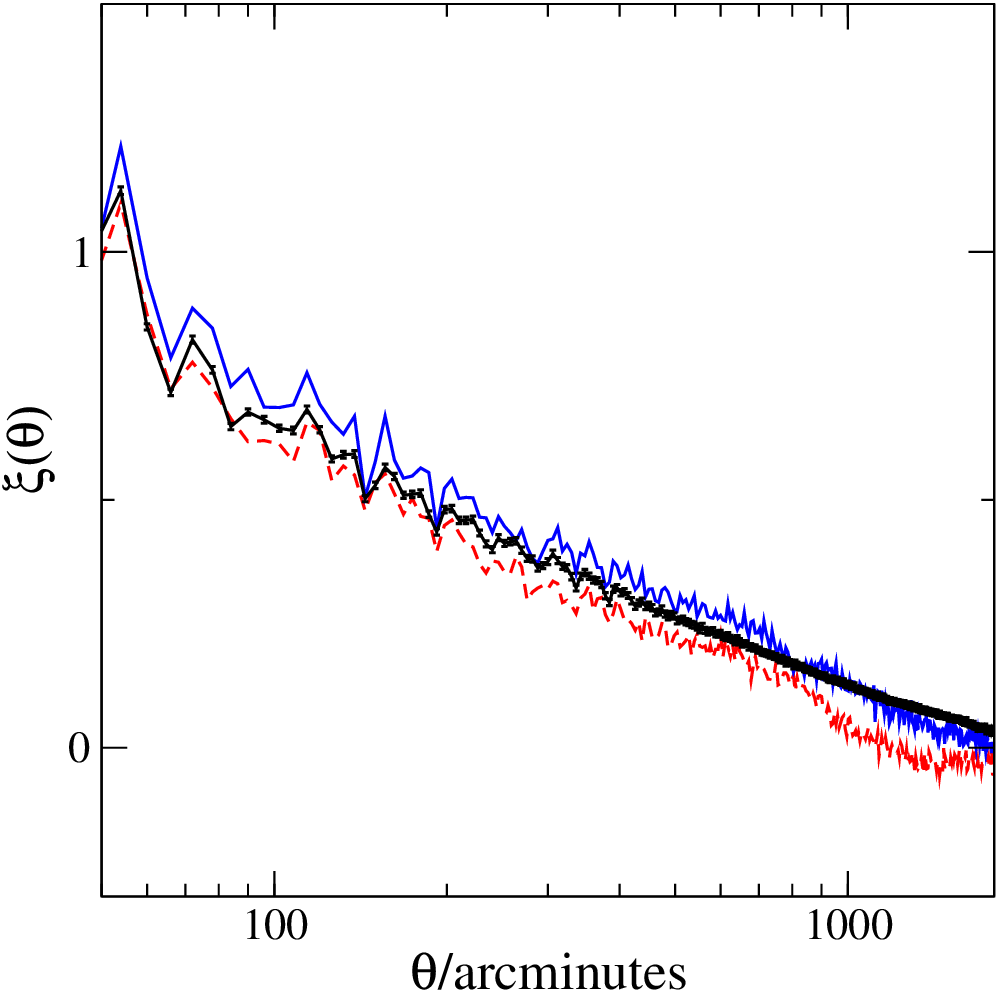}
\caption{\small The peak-peak correlation function of WMAP's data hot spots in the
dashed (red)
line and 
cold spots in the solid (blue) line. Simulated data (averaged over
100 Gaussian simulations) in the middle (black) line -  the error bars shown are
the errors on the mean. Note that as these are correlation functions,
the errors are correlated.} 
\label{fig:full_sky}
\end{center}
\end{figure}

\begin{figure}
\vspace{0.6cm}
\begin{center}
\includegraphics[width=2.in]{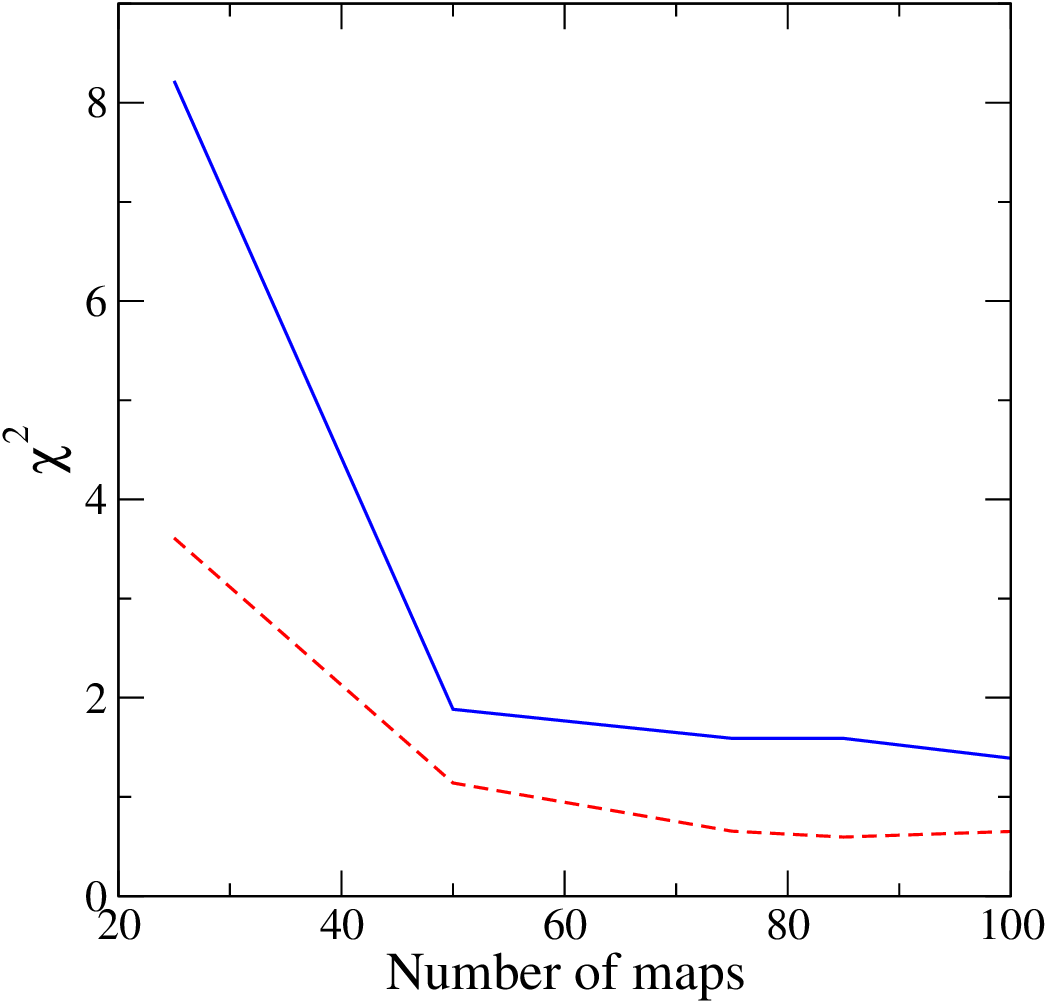}
\caption{\small Convergence of $\xi_C$ (solid, blue line) and $\xi_H$
(dashed, red line) with number of Gaussian maps
used to estimate $\bar\xi^G_i$ and $C_{ij}$ as defined in Section 3.} 
\label{fig:convergence_fs}
\end{center}
\end{figure}

We first consider all the hot spots above a certain threshold
$\nu\sigma$ (or cold spots below a negative threshold $-\nu\sigma$),
for the entire sky except for the masked regions of galactic plane
and point sources. The results for a threshold of $\nu=1.5$
are shown in Figure \ref{fig:full_sky}. However, with no specific type of
non-Gaussianity in mind there is no particular
reason to choose any given threshold - too high a threshold reduces
the number of peaks and the study is limited by cosmic variance, too
low and the computation time is uncomfortably long. Our choice was a
compromise to obtain a large number of extrema to get small errors,
whilst not being computationally too expensive. We also plot
the peak-peak correlation function averaged over 100 Gaussian maps and
the error bars on the Gaussian curve are the errors on the mean. The small error
bars show good convergence of the average of the peak-peak correlation
function from the 100 Gaussian maps. Figure
\ref{fig:convergence_fs} shows the convergence of $\xi_H$ and $\xi_C$
with increasing number of maps.

Although not optimally  sampled, the structure we see at small angular scales is real structure, as
expected from 
\citet{art:HeavensSheth99} and \citet{art:HeavensGupta01}

\begin{figure}
\vspace{0.6cm}
\begin{center}
\includegraphics[width=2.3in]{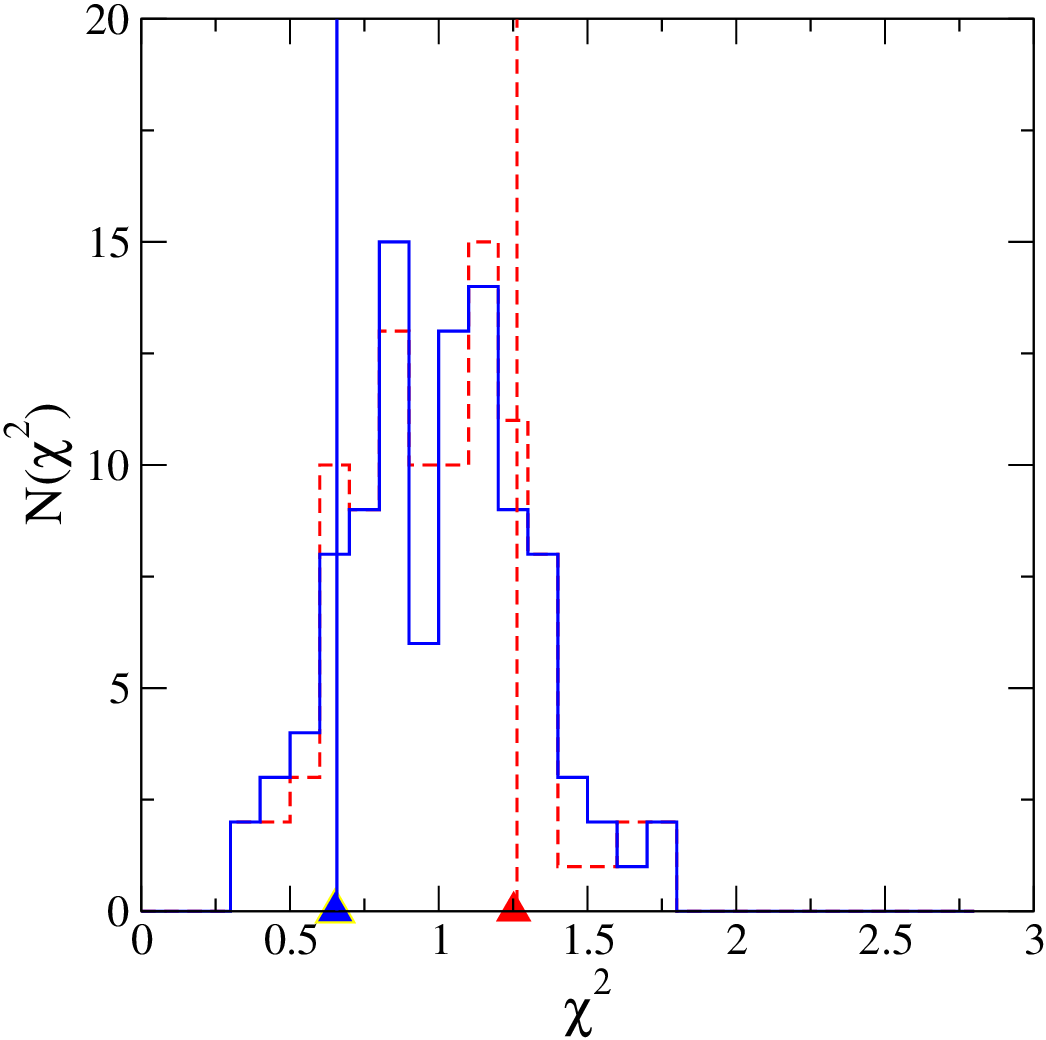}
\caption{\small The distribution of reduced $\chi^2$ values for all
of the 100 Gaussian maps: hot spots in the dashed (red) line and cold spots in the solid (blue) line. The $\chi^2$ values for the WMAP data are represented by the small
triangles and vertical lines.} 
\label{fig:full_sky_chi}
\end{center}
\end{figure}

We see immediately that neither the hot spots nor the cold spots follow
the Gaussian simulations - the cold spots show an excess of correlation
whereas the hot spots show a lack of correlation with respect to the
Gaussian simulations. 
These differences are, however,
not significant; one disadvantage of the correlation function is that
the errors can be highly correlated. The distribution of the $\chi^2$ values for all
of the Gaussian maps can be seen in Figure \ref{fig:full_sky_chi}, together with the
values for the WMAP data.
We find $\chi^2=1.25$ for the hot spots and $\chi^2 = 0.65$ 
for the cold spots, which are within the Gaussian $1 \sigma$
confidence level. So
the maps analysed in this way do not show any sign of non-Gaussianity. This is
in agreement with \citet{art:LarsonAndWandelt05} who also find no significant deviation
from Gaussianity when they compute the peak-peak correlation of hot
and cold spots (although they work with lower resolution maps).

There have been numerous claims in the literature (see Introduction) that the
WMAP maps show an asymmetry in their statistical properties 
between the Northern and the Southern hemispheres, so we turn to this next.

\subsection{North-South analysis}

To further investigate any discrepancy between the WMAP data and our
Gaussian simulations we apply the masks 30BN and 30BS to all our
maps and proceed to get an estimation of the peak-peak correlation
function for the Northern and Southern hemispheres separately.

\begin{figure}
\vspace{0.6cm}
\begin{center}
\includegraphics[width=2.3in]{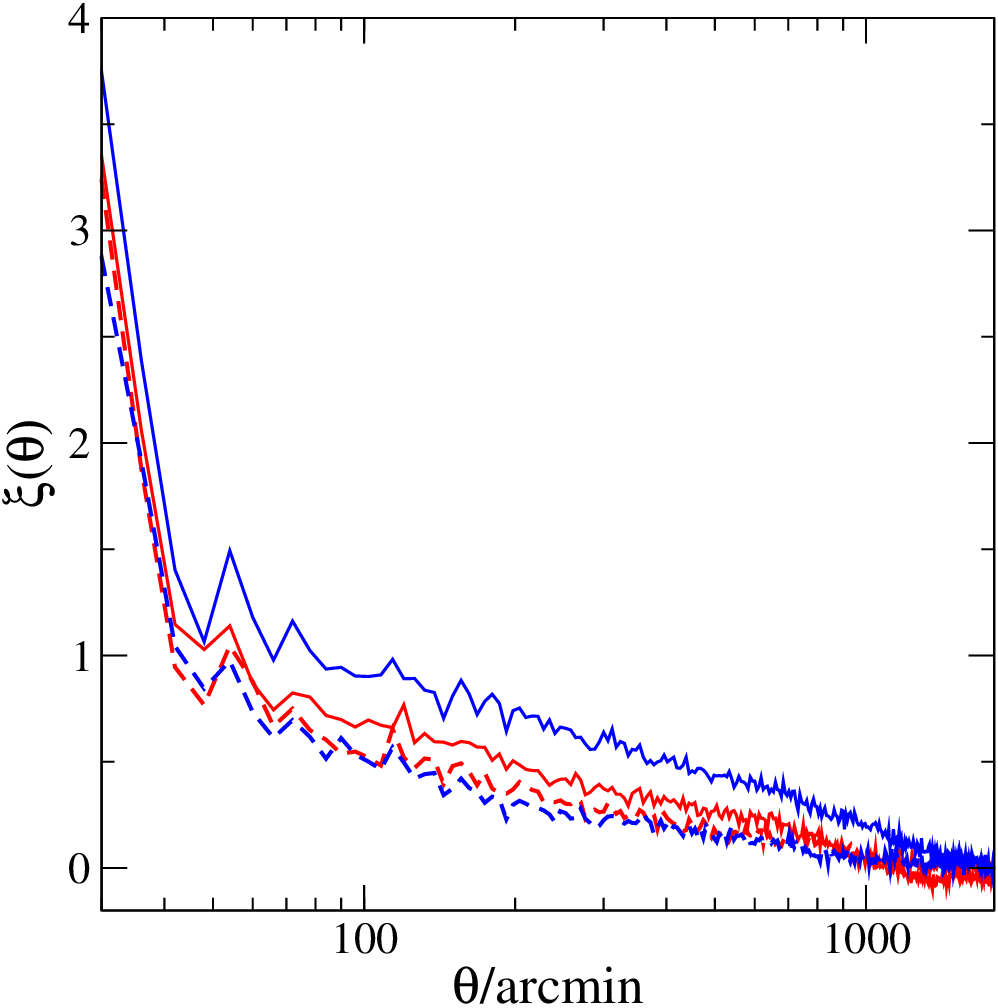}
\caption{\small The peak-peak correlation for the WMAP data in the two
hemispheres - solid lines show the South and dashed lines the
North. The inner pale (red) lines show hot spots and the outter (blue)
lines show cold spots.}
\label{fig:NS_corr_HPF0}
\end{center}
\end{figure}

Figure \ref{fig:NS_corr_HPF0} shows the peak-peak correlation function
of the WMAP data for cold and
hot spots calculated in the Northern and Southern hemispheres.
We find a difference between the correlation of cold spots in the
different hemispheres. Again we use a $\chi^2$ statistic,
$\chi^2_{NS}$ for $\Delta\xi_C$ and $\Delta\xi_H$, with the mean and covariance matrix
estimated from 250 Gaussian maps. By analysing each hemisphere
seperately, we are reducing the number of peaks available for the
estimation of the peak-peak correlation function. Hence we found that a
greater number of maps was needed to ensure good convergence of the
average peak-peak correlation function and of the covariance
matrix. See Figure \ref{fig:convergence} in the next section for convergence of some of
the statistics with increasing number of maps.

\begin{figure}
\vspace{0.6cm}
\begin{center}
\includegraphics[width=2.3in]{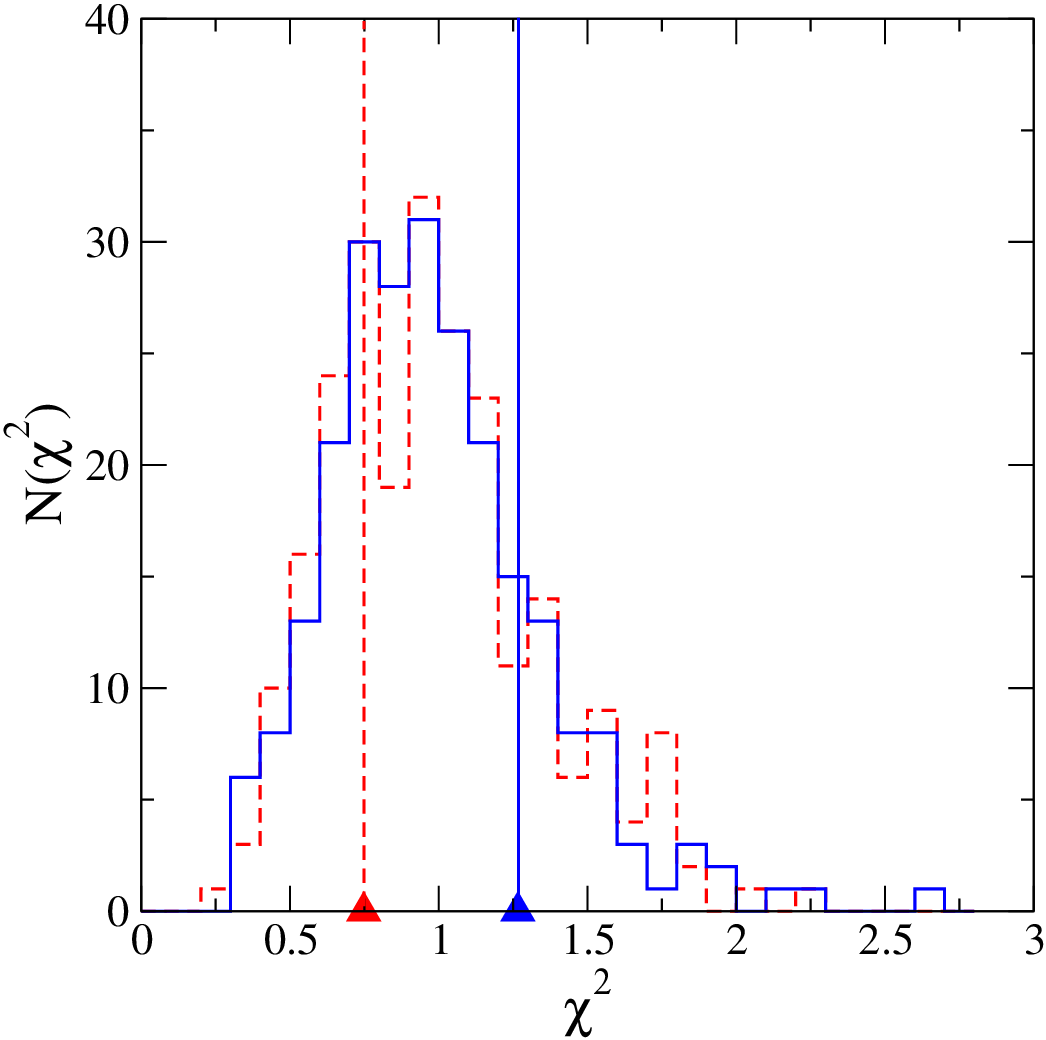}
\caption{\small The distribution of reduced $\chi^2_{NS}$ for all 250 Gaussian
maps. Hot spots in the dashed (red) line and cold spots in the solid (blue) line. The $\chi^2_{NS}$ values
for the WMAP data are represented by the small triangles and vertical lines. } 
\label{fig:chi_dist_HPF0}
\end{center}
\end{figure}

We calculate $\chi^2_{NS}$ for our ensemble of
Gaussian maps, whose distribution can be seen in Figure \ref{fig:chi_dist_HPF0}, together
with the $\chi^2_{NS}$ value calculated for the WMAP data, for hot and
cold spots.

We note that the fact we are finding the South-North difference not to be
significant ($\chi^2_{NS} = 0.748$ and $1.27$ for hot and cold spots
respectively) may be due to the fact that the peak-peak correlation
function of threshold-selected peaks is highly sensitive to cosmic
variance in the low multipoles. All the estimators are highly
correlated and are shifted up and down in synchrony from Gaussian
realisation to Gaussian realisation: the noisy low-$\ell$ multipoles
can shift large numbers of peaks above or below the threshold
depending on the mode amplitude.   This suggests that the use of a
high-pass filter - effectively removing the signal from cosmic
variance above a given angular scale on
the sky - may be an efficient way to increase the sensitivity
to non-Gaussian features.

\subsection{Constraining in harmonic space}

We high-pass filter our maps (real and Gaussian) by
constructing several window functions 
given by ${\mathcal W}_{\ell_{cut}}(\ell) = 0$ for $\ell
\le \ell_{cut}$, ${\mathcal W}_{\ell_{cut}}(\ell) = 1$ otherwise,  which we apply to our maps and we investigate cases
with $\ell_{cut} = 0, 5, 10, 15, 20, 25, 30$ and $40$. \\

We mask the WMAP maps before filtering. This is necessary
because of the presence of foregrounds - the strong ringing effect
in pixel space which results from such a sharp cut-off in harmonic
space causes unwanted foreground signal to leak from the masked region. We
follow the algorithm described below:

\begin{itemize}
\item We mask the WMAP data.
\item We extract the $a_{lm}$ coefficients using HEALPix's
routine \verb,anafast,.
\item We multiply the harmonic coefficients by the respective window
function.
\item We generate the map in pixel space by using the HEALPix's
routine \verb,synfast,.
\item We re-mask the map. 
\item We remove the monopole/dipole from the unmasked regions.
\end{itemize}

Since there is no foreground contamination in the Gaussian maps, there is no need to apply the initial mask.
For testing purposes, we applied both methods to a number of Gaussian
maps and found them to produce the same final results.

For each case
we construct an ensemble of 250 Gaussian maps. We compute $\Delta\xi_H$, $\Delta\xi_C$,
$\xi_{NH}$, $\xi_{NC}$, $\xi_{SH}$, $\xi_{SC}$, $\tilde{\xi}_H$ and
$\tilde{\xi}_C$ 
for all our Gaussian maps as well as the WMAP data.
Figure \ref{fig:convergence} shows convergence of $\xi_{SC}$ and
$\Delta\xi_C$ with number of maps in the
solid and dashed lines respectively. The same plot also shows the
convergence of the same statistics but this time calculated in a
single-frequency Q-band map (see Section 4.5 for a single-frequency analysis).
We quote all significance levels in Table \ref{tab:summary}. 

\begin{figure}
\vspace{0.6cm}
\begin{center}
\includegraphics[width=2.3in]{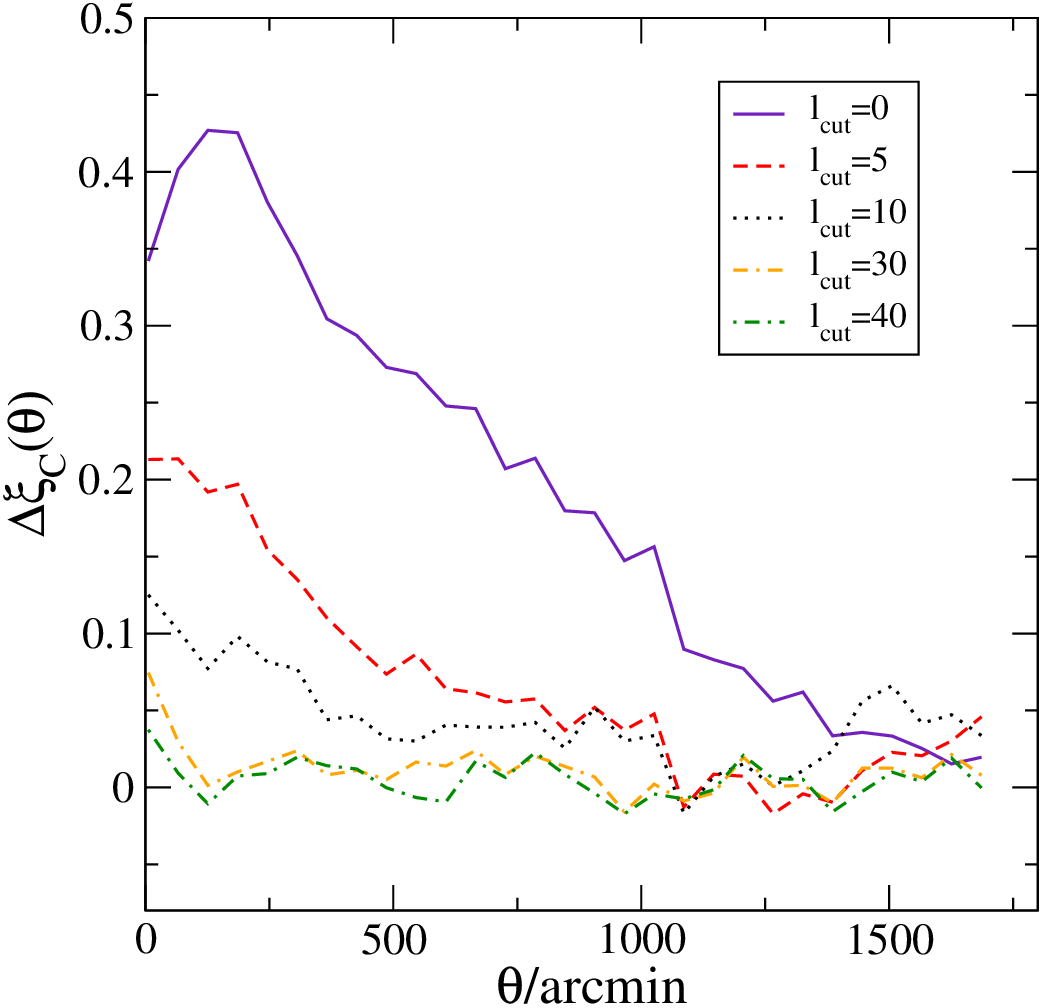}
\caption{\small $\Delta\xi_C(\theta)$ for the WMAP map high-pass
filtered with different values of $\ell_{cut}$.} 
\label{fig:lmin_cold}
\end{center}
\end{figure}

Figure \ref{fig:lmin_cold} shows $\Delta\xi_{C}(\theta)$ 
for some different $\ell_{cut}$ in the WMAP data. 
We note that the difference between the Southern and Northern
hemispheres decreases as we remove more and more of the low order
multipoles. This could be either due to the fact that cosmic variance alone
is to blame for the North/South difference we see, or it could be due
to the fact that whatever is causing this North/South difference is
intrinsically a large scale effect. 

\begin{figure}
\vspace{0.6cm}
\begin{center}
\includegraphics[width=2.3in]{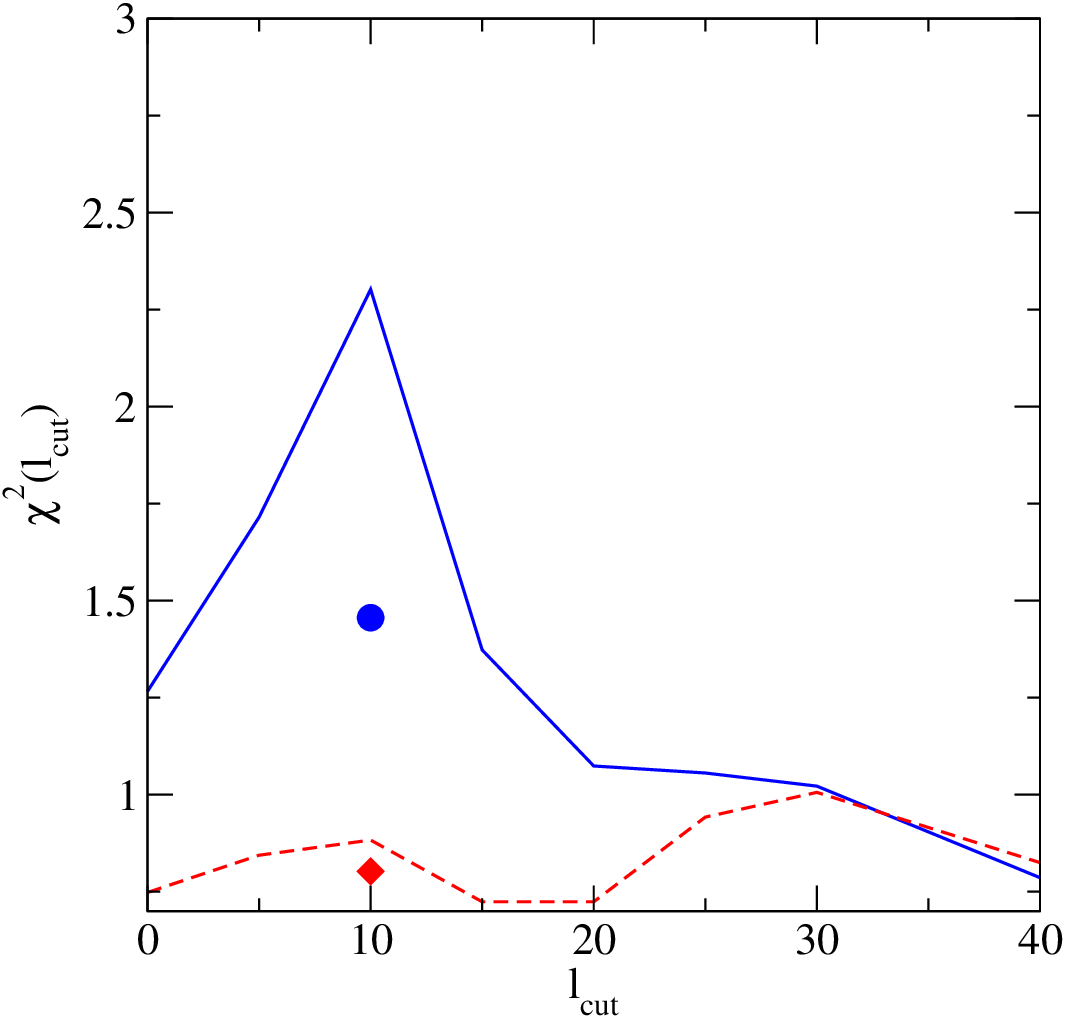}
\caption{\small $\chi^2_{NS}$ as a function of $\ell_{cut}$ for the
WMAP data. Hot spots
in the solid (red) line, cold spots in the dashed
line (blue). The circle (blue) and the diamond (red)
are the $\chi^2$ value (cold spots and hot spots respectively) for runs with the regions
of sky within $30$ degrees of the galactic plane removed (see Section
\ref{sec:real_space}). We recall that the 99$\%$ confidence levels for
a $\chi^2$ distribution of 18 degrees of freedom come at $\chi^2 \le
0.42$ and $\chi^2 \ge 1.93$. See Figure \ref{fig:added_hist_NS} for
the distribution of the $\chi^2_{NS}$ values of all Gaussian maps.} 
\label{fig:chi_NS_lmin}
\end{center}
\end{figure}

\begin{figure}
\vspace{0.6cm}
\begin{center}
\includegraphics[width=2.3in]{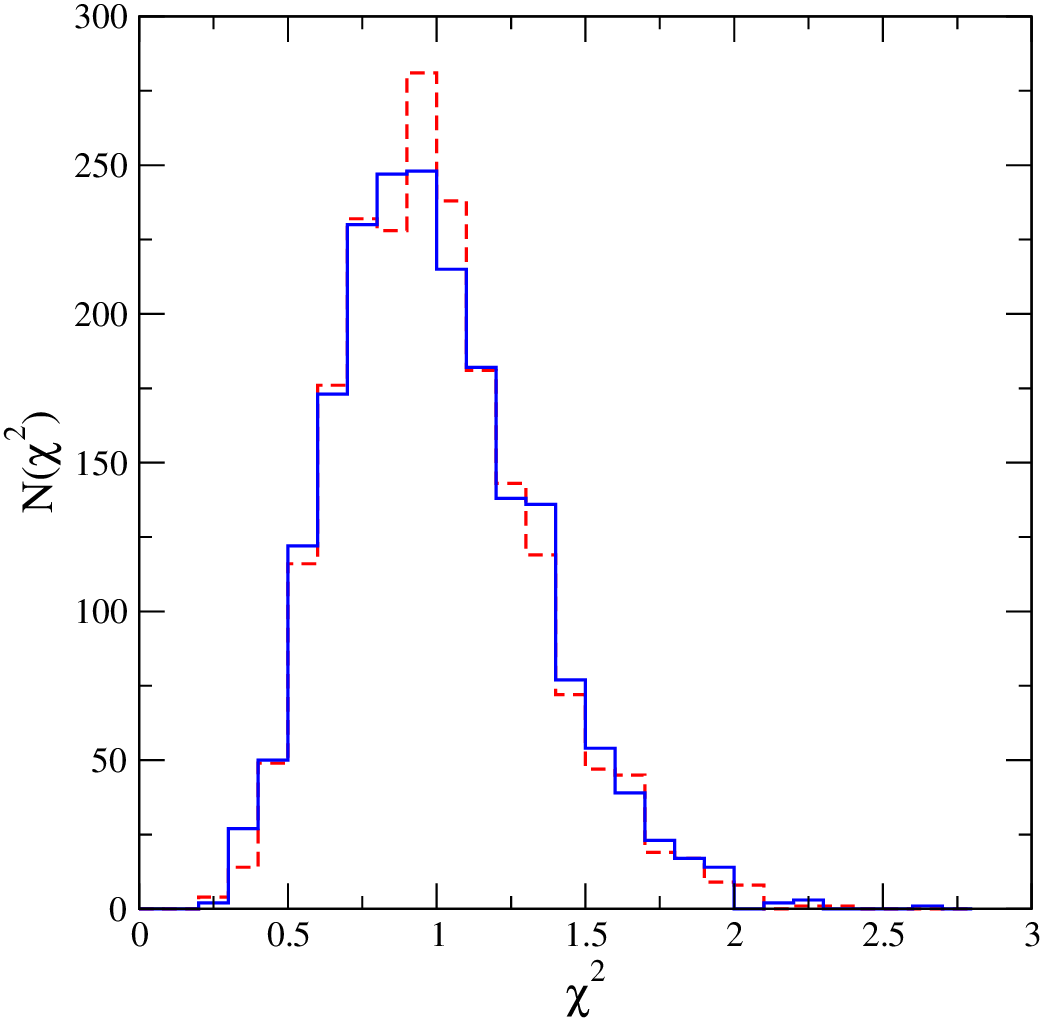}
\caption{\small The added distribution of $\chi_{NS}^2$ values for all our
Gaussian maps at all different $\ell_{cut}$. Hot spots in the dashed
(red) line, cold spots in the solid (blue) line. Similar histograms
were produced for all of our other statistics, and all show an
identical added distribution of reduced $\chi^2$ values. } 
\label{fig:added_hist_NS}
\end{center}
\end{figure}

We test the significance of each of these differences by
using $\chi^2_{NS}$. Figure \ref{fig:chi_NS_lmin}
shows $\chi_{NS}^2(\ell_{cut})$ for cold and hot spots. We plot
the distribution of $\chi^2_{NS}$ using {\it all} the different
$\ell_{cut}$ Gaussian maps - these maps are not strictly independent 
(although the statistics share the same underlying $\chi^2$ distribution over all values
of $\ell_{cut}$) so we use only the 250 independent maps at each $\ell_{cut}$
to draw conclusions about the significance of each detection - see
Section \ref{sec:summary}. The added histogram over the 2000 maps can
be seen in Figure
\ref{fig:added_hist_NS}.

\begin{figure}
\vspace{0.6cm}
\begin{center}
\includegraphics[width=2.3in]{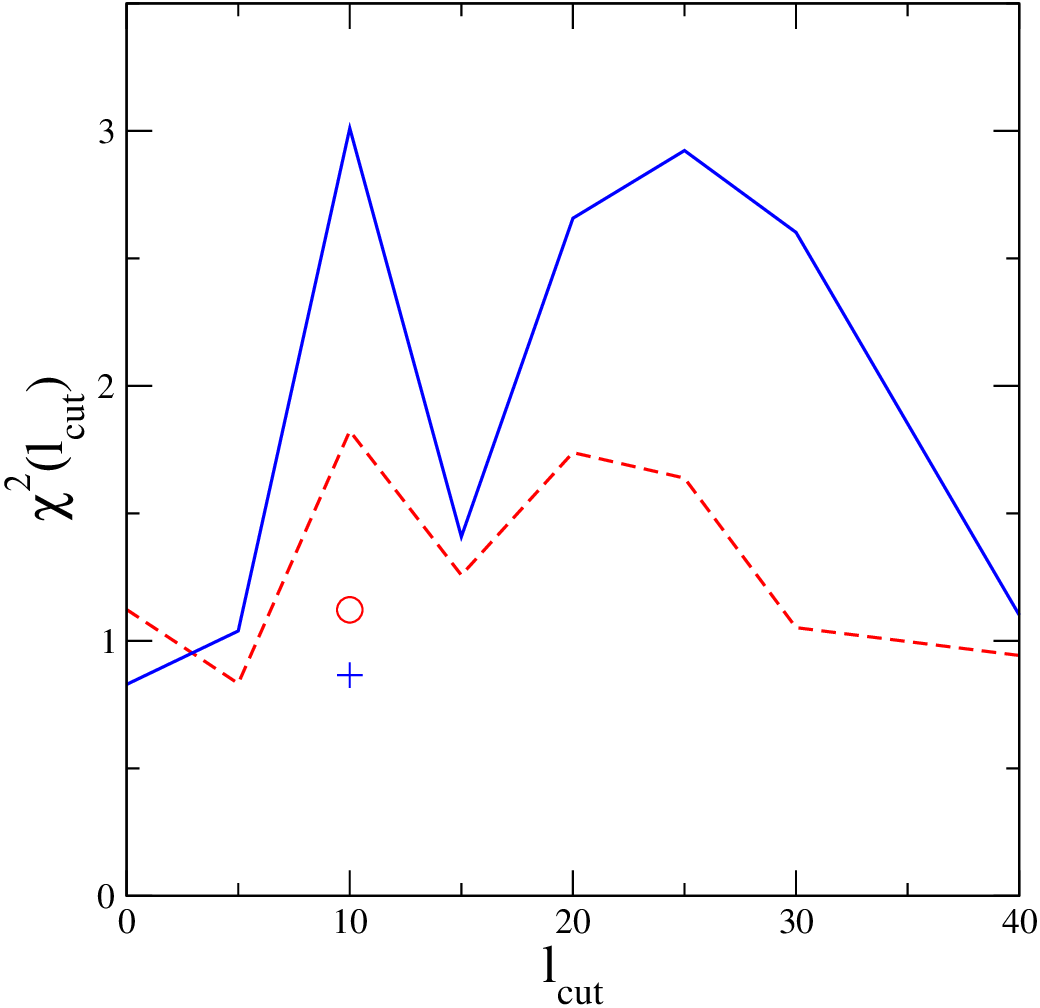}
\caption{\small $\chi^2(\ell_{cut})$ for $\tilde{\xi}_C$ solid (blue)
line and $\tilde{\xi}_C$ dashed (red) line for the WMAP data. The single points at
$\ell_{cut}=10$ are the $\chi^2$ values for cold spots (blue cross) and
hot spots (red circle) in runs with the regions of
sky within $30$ degrees of the galactic plane removed (see Section
\ref{sec:real_space}).}
\label{fig:chi_lmin_fs}
\end{center}
\end{figure}

\begin{figure}
\vspace{0.6cm}
\begin{center}
\includegraphics[width=2.3in]{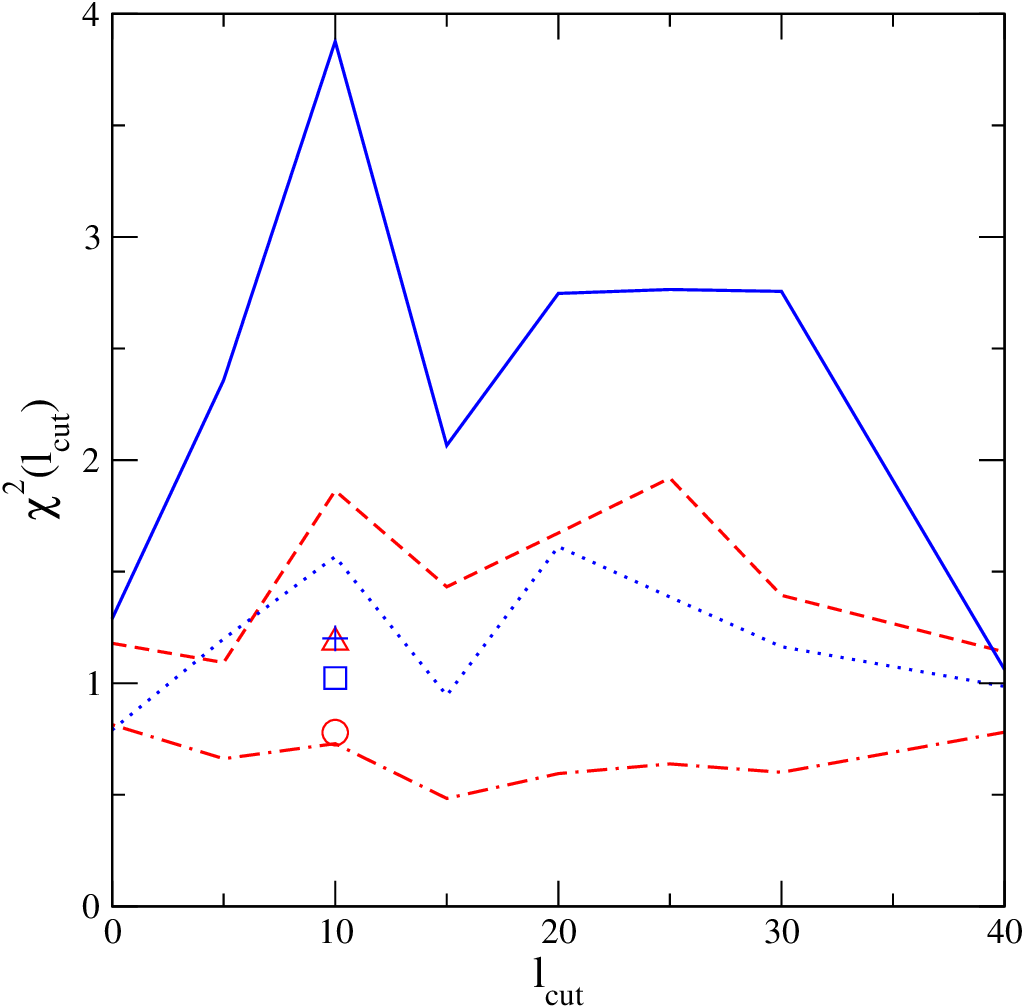}
\caption{\small $\chi^2(\ell_{cut})$ for $\xi_{NC}$ (blue dotted
line), $\xi_{NH}$ (red dot-dashed line), $\xi_{SC}$ (blue solid line)
and $\xi_{SH}$ (red dashed line) for the WMAP data. The single point at $\ell_{cut}=10$
are the $\chi^2$ values for runs with the regions of sky within $30$
degrees of the galactic plane removed (see Section
\ref{sec:real_space}): $\xi_{NH}$ in the red circle, $\xi_{NC}$ in the
blue square, $\xi_{SH}$ in the red triangle and $\xi_{SC}$ in the blue
cross.}
\label{fig:chi_indv_lmin}
\end{center}
\end{figure}

\begin{figure}
\vspace{0.6cm}
\begin{center}
\includegraphics[width=2.3in]{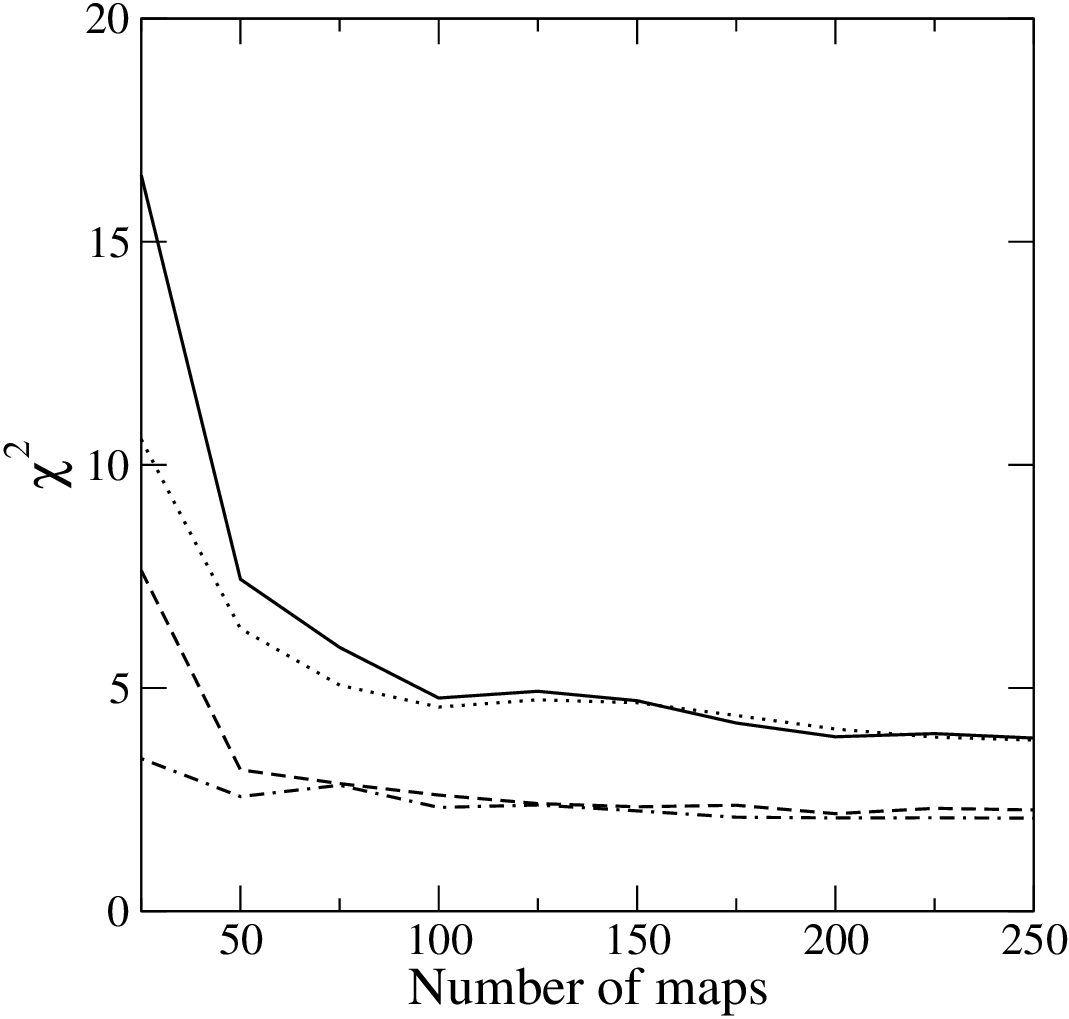}
\caption{\small Convergence of some of our statistics which yielded
detections of non-Gaussianity with increasing number of maps. For the QVW map we show $\xi_{SC}$ in
the solid line and $\Delta\xi_C$ in the dashed line. For the
single-frequency Q-band map we show $\xi_{SC}$ in the dotted line and
$\Delta\xi_C$ in the dot-dashed line}
\label{fig:convergence}
\end{center}
\end{figure}

We do the same test and construct identical plots for all our statistics:
$\tilde{\xi}_H$/$\tilde{\xi}_C$ in Figure \ref{fig:chi_lmin_fs} 
and $\xi_{NH}$/$\xi_{NC}$/$\xi_{SH}$/$\xi_{SC}$ in Figure
\ref{fig:chi_indv_lmin}. The added histograms across all values of $\ell_{cut}$ for these
statistics are very similar to that shown in Figure
\ref{fig:added_hist_NS}.

The first point
to make is that the non-Gaussianity is consistently absent at
$\ell_{cut}=40$: there is no evidence from the peak-peak correlation function of non-Gaussianity on scales with $\ell > 40$.

The most significant non-Gaussian detections come from
the cold spots in the Southern hemisphere, $\xi_{SC}$, 
at $\ell_{cut}=10$ (with a value of $\chi^2=3.877$ and a probability formally estimated to be
$10^{-7}$ - see Section \ref{sec:summary}), where
we also find significant detections in the South-North difference for
cold spots, $\Delta\xi_C$ $(\chi^2_{NS}=2.302)$, and in the average of Northern and Southern hemispheres
for cold spots, $\tilde{\xi}_C$ $(\chi^2=3.011)$.
In addition to this, we have less significant detections at
$\ell_{cut}=20, 25$ and $30$ in $\xi_{SC}$ ($\chi^2=2.747, 2.764$ and
$2.756$ respectively) and $\tilde{\xi}_C$ ($\chi^2 = 2.658, 2.923$ and
$2.601$, see
Figures \ref{fig:chi_lmin_fs} and \ref{fig:chi_indv_lmin}). All of these
do not appear in a North minus South analysis. This could be
simply because the signal is not significant enough to show up in such
analysis (we are roughly doubling the variance of our estimator by
subtracing the data of the Sorthern and Northern hemispheres).

\subsection{Constraining in real space}
\label{sec:real_space}
We further investigate the origin of this detection
by removing extra regions near the
masked galactic plane. We work on the maps where the significance of
the signal is the strongest (those with $\ell_{cut}=10$), which we
mask with the extended masks: bcut30\_knorth and bcut30\_ksouth.

We proceed the same way as before and compute the full set of
estimators:  $\Delta\xi_H$, $\Delta\xi_C$,
$\xi_{NH}$, $\xi_{NC}$, $\xi_{SH}$, $\xi_{SC}$, $\tilde{\xi}_H$ and
$\tilde{\xi}_C$ for all our Gaussian maps as well as the WMAP data and
use the adequate $\chi^2$ statistic for each of them to test the WMAP
data for non-Gaussianity (we generate new random catalogues whose
spacial distribution follows that of the new masks).

Figures \ref{fig:chi_NS_lmin}, \ref{fig:chi_lmin_fs} and \ref{fig:chi_indv_lmin} show how the new $\chi^2$ values compare with the ones
previously obtained when we did not use any extra galactic cut - all 
values drop significantly to values which are perfectly consistent
with the Gaussian hypothesis (the most extreme value being
$\chi^2_{NS}=1.46$ for $\Delta\xi_C$) , indicating that our significant non-Gaussian
detection in the cold spots is located within $30$ degrees of the
galactic plane. Clearly, this hints at residual foreground
contamination associated with the Milky Way.

We note that we have only tested this on maps
with $\ell_{cut}=10$ since this is where we have found our strongest
detection. We cannot discard the possibility that the
effect that yields detections on maps with $\ell_{cut}=15, 25$ and
$30$ is a different effect altogether which does not lie in the
galactic region.

\subsection{A single-frequency analysis}\label{sec:single_freq}

To check whether the non-Gaussian signal we detect is related to possible
residual foregrounds in the WMAP data we conduct a single frequency
analysis of the maps. Indeed, the expected galactic foreground
contribution to the WMAP maps consists mainly of
synchrotron, free-free and dust emission. All of these effects are
frequency-dependent and obviously non-Gaussian. 
If any foreground residuals are still present in
the foreground-cleaned data then we would expect them to 
contribute differently to each of the different frequency maps. 
We note that any
residual noise may also contribute differently to each frequency.

We construct the real map and each of the 250 simulated single 
frequency maps by following the same
method we used to construct the co-added QVW map (described in Section
\ref{sec:maps}), 
but we use only the relevant DAs for each of the frequencies. We then
smooth the WMAP and Gaussian maps with a 12 arc minute FWHM Gaussian 
beam and high-pass filter with a $\ell_{cut}=10$ window function
(where we had the most significant non-Gaussian detection).

\begin{figure}
\vspace{0.6cm}
\begin{center}
\includegraphics[width=2.3in]{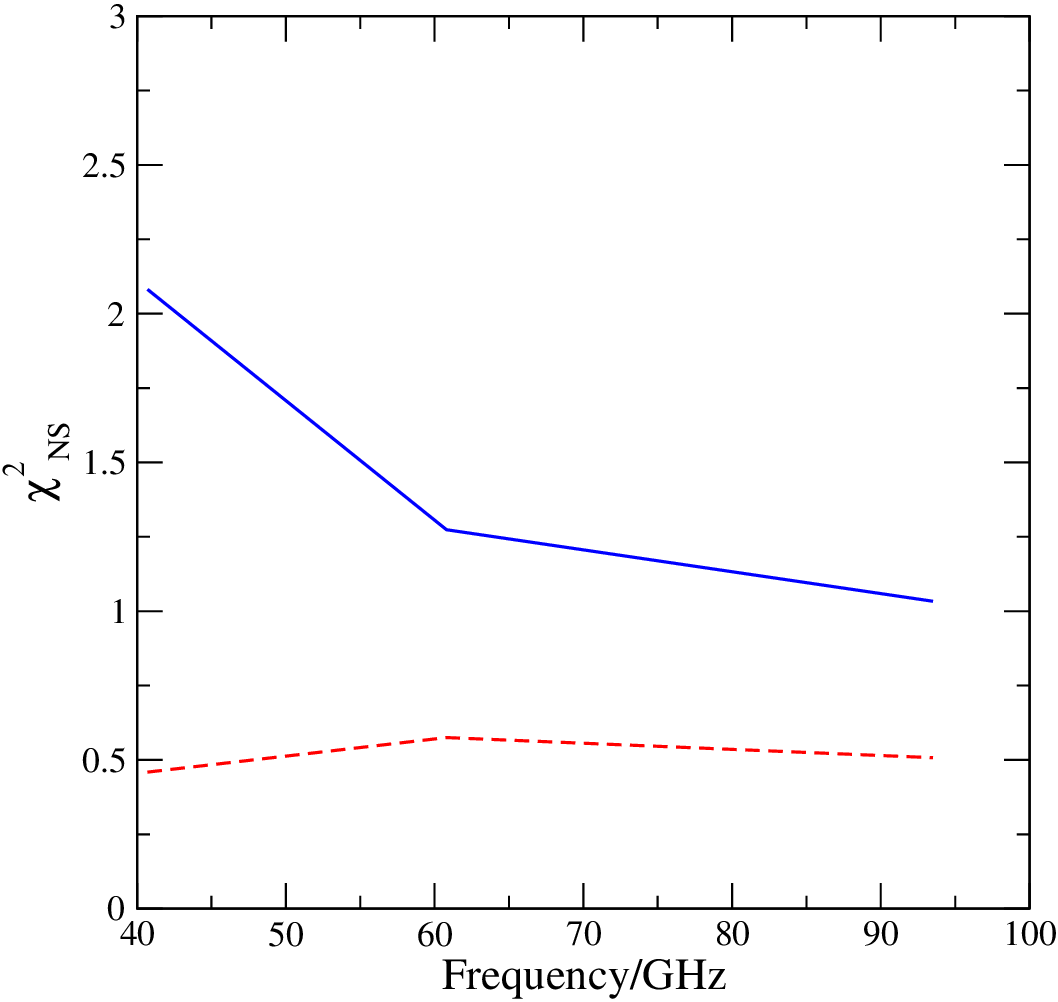}
\caption{\small $\chi^2_{NS}$ for all three frequencies: Q (41 GHz), V
(61 GHz) and W (94 GHz) on maps with $\ell_{cut}=10$. Statistics for cold
spots in the solid (blue) line, for
hot spots in the dashed (red) line.} 
\label{fig:chi_freq_NS}
\end{center}
\end{figure}

\begin{figure}
\vspace{0.6cm}
\begin{center}
\includegraphics[width=2.3in]{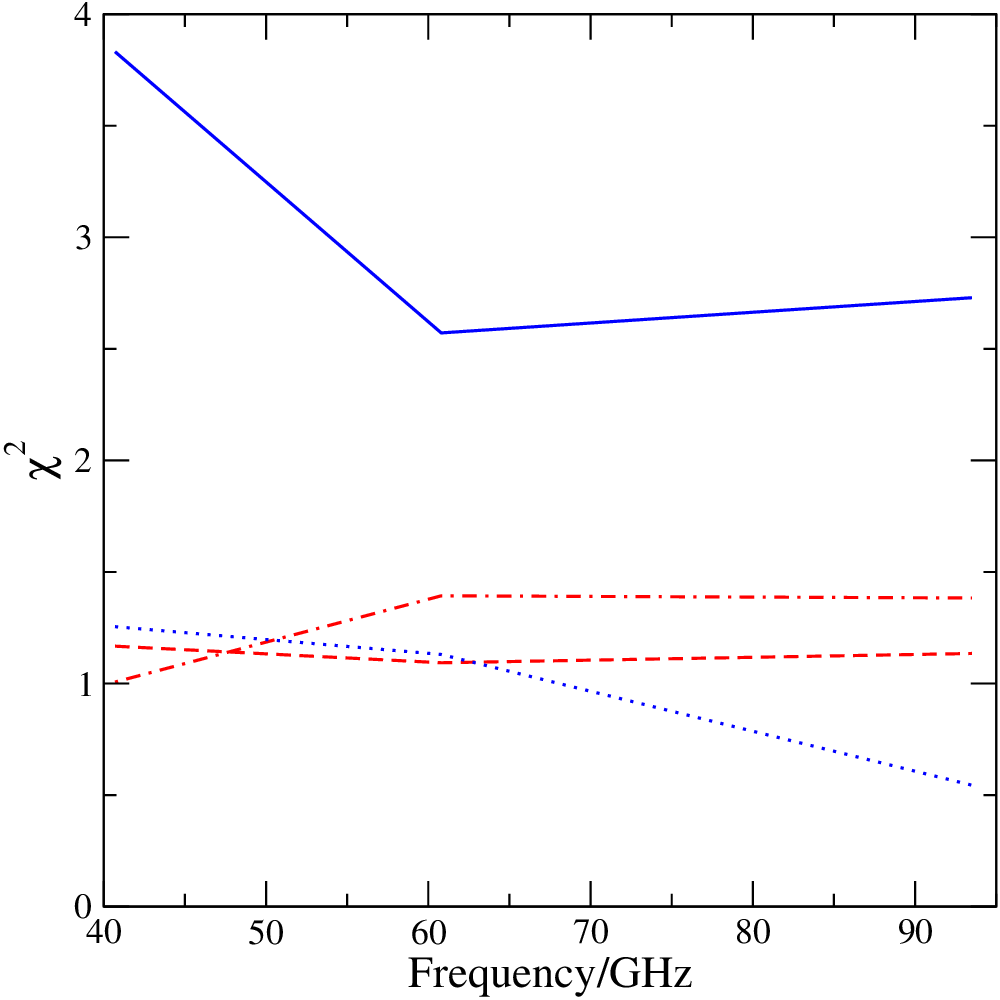}
\caption{\small $\chi^2$ for all three frequencies: Q (41 GHz), V
(61 GHz) and W (94 GHz) on maps with $\ell_{cut}=10$. Statistics for
$\xi_{NH}$ in the dot dashed (red) line, $\xi_{NC}$ in the dotted (blue)
line, $\xi_{SH}$ in the dashed (red) line and finally for $\xi_{SC}$ in
the solid (blue) line.} 
\label{fig:all_chi_freq}
\end{center}
\end{figure}

\begin{figure}
\vspace{0.6cm}
\begin{center}
\includegraphics[width=2.3in]{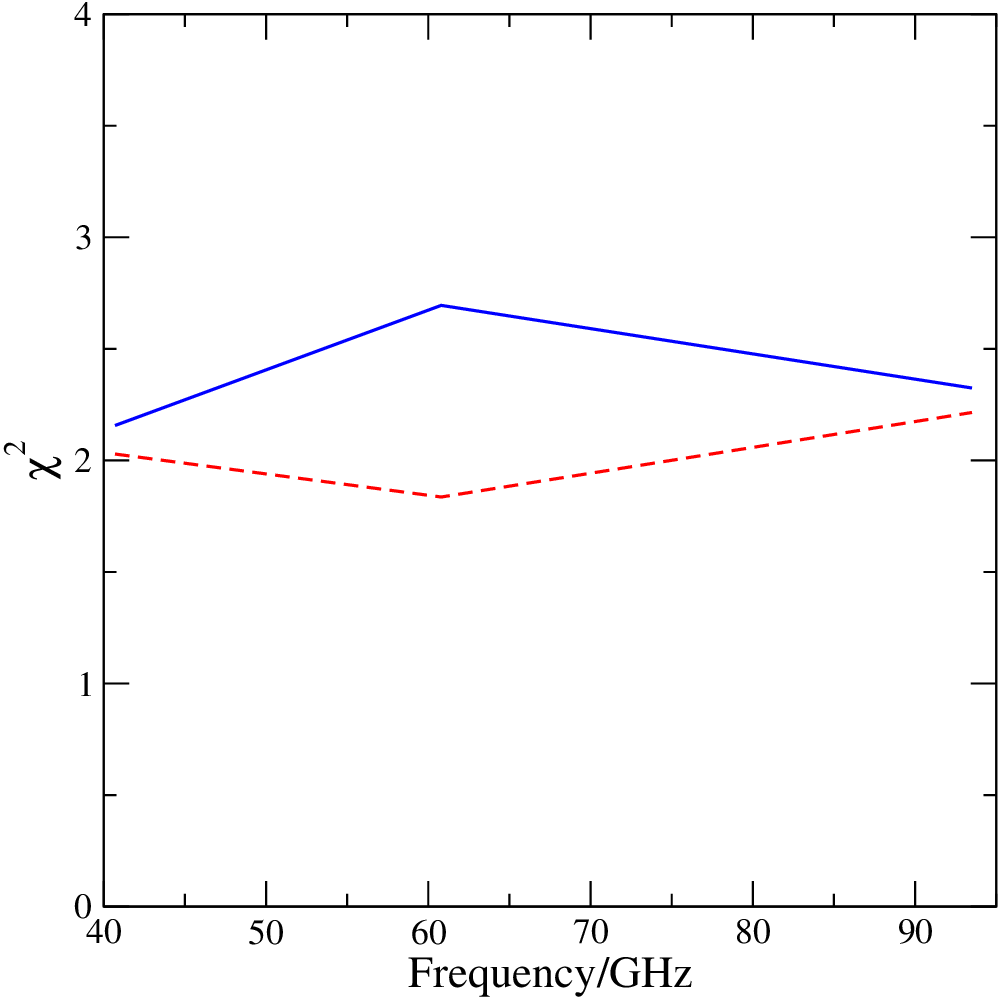}
\caption{\small $\chi^2$ for all three frequencies: Q (41 GHz), V
(61 GHz) and W (94 GHz) on maps with $\ell_{cut}=10$. Statistics for
$\tilde{\xi}_H$ in the dashed (red) line and for $\tilde{\xi}_C$ in the
solid (blue) line.} 
\label{fig:chi_fs_freq}
\end{center}
\end{figure}

We calculate the full set of estimators for each of the
frequencies: $\Delta\xi_H$, $\Delta\xi_C$,
$\xi_{NH}$, $\xi_{NC}$, $\xi_{SH}$, $\xi_{SC}$, $\tilde{\xi}_H$ and $\tilde{\xi}_C$, for which the $\chi^2$ values can be seen in Figures
\ref{fig:chi_freq_NS}, \ref{fig:all_chi_freq} and \ref{fig:chi_fs_freq}.

We find significant non-Gaussian signals coming from
the cold spots $\Delta\xi_C$ in the Q band ($\chi^2_{NS}=2.081$) and $\xi_{SC}$ in all three bands,
although it is strongest in the Q band (with $\chi^2 = 3.831$). We also find detections
in our full-sky estimates in the cold spots in all three bands, and,
for the first time, in the hot spots in bands Q and W (see Figure \ref{fig:chi_fs_freq}).

We may be seeing a
frequency-dependent type of non-Gaussianity, although we can not put
aside the possibility of a cosmological origin. To improve
readability we do not present the plots with the $\chi^2$
distributions of the 250 Gaussian maps for each of the frequencies and
for each of the estimators. We do, however, quote the number of
Gaussian maps with a $\chi^2_{Gaussian} \ge \chi^2_{WMAP}$ for all
significant detections in Table \ref{tab:summary}, Section \ref{sec:summary}.

\subsection{Removing the cosmological signal}	
\label{sec:differences}
In order to investigate the possibility of any contributions from
foregrounds or unexplained noise properties, we remove what is
taken to be the cosmological signal from our analysis.

To do so we subtract different single-frequency maps to produce three maps which
contain only a mix of subtracted residual foregrounds (if any) and noise. We produce a
$V-Q$, a $V-W$ and a $Q-W$ map, which are simply a pixel-by-pixel
subtraction of each of the single frequency maps, constructed as
described in Section \ref{sec:single_freq}.

With the cosmological signal removed, the detailed noise properties of
these 3 subtracted maps at large angular scales now become
important for our analysis and one should be careful when
constructing equivalent Gaussian maps (see Section 2.1.2).
We therefore take a slightly different route to construct
the Gaussian simulations with which we compare the WMAP data, and we
now make use of the 110 noise simulations supplied by the WMAP team.
We construct single-frequency noise maps by adding the respective
individual radiometer simulations following the same weighting scheme as
described in Section \ref{sec:maps}, which we then smooth and
high-pass filter with a $\ell_{cut}=10$ window. We then subtract
different frequency noise maps in order to produce 110 simulations
with which we compare our real $V-Q$, $V-W$ and $Q-W$ maps. We reemphasize that for maps which include the signal, the non-white nature of the noise at low-$\ell$ is essentially irrelevant, as the signal dominates entirely ~\citep{art:HinshawEtAl03}.

We construct $\Delta\xi_H$, $\Delta\xi_C$,
$\xi_{NH}$, $\xi_{NC}$, $\xi_{SH}$, $\xi_{SC}$, $\tilde{\xi}_H$ and
$\tilde{\xi}_C$ for the simulations and the real data as before and use the
respective $\chi^2$ statistic to probe for non-Gaussian signatures.

\begin{figure}
\vspace{0.6cm}
\begin{center}
\includegraphics[width=2.3in]{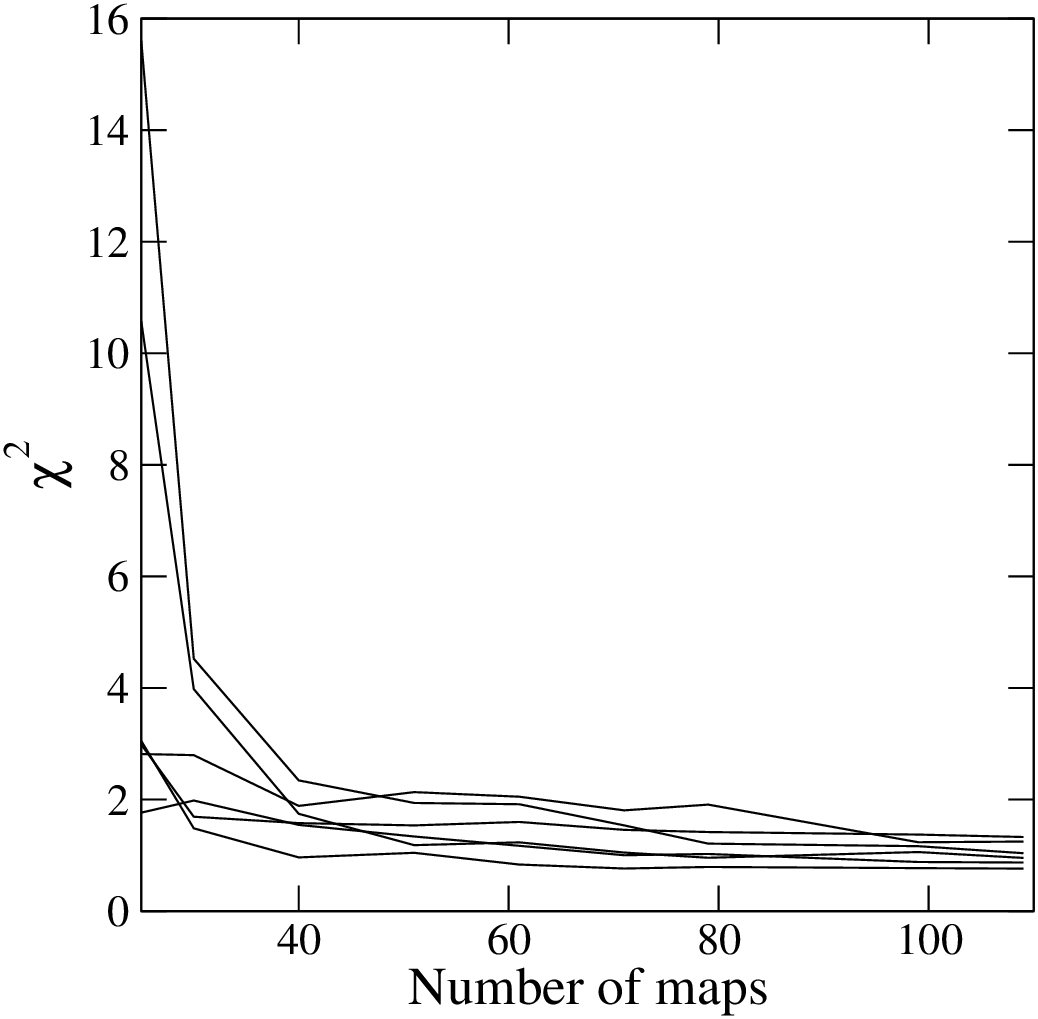}
\caption{\small Convergence of the reduced $\chi^2$ values for $\Delta\xi_H$, $\Delta\xi_C$,
$\xi_{NH}$, $\xi_{NC}$, $\xi_{SH}$ and $\xi_{SC}$ in the $Q-W$ map as
a function of number of Gaussian maps.}
\label{fig:convergence_QW}
\end{center}
\end{figure}

In this case, our total number of maps was constrained by the number of noise
simulations provided by the WMAP team. Figure \ref{fig:convergence_QW}
shows how the reduced $\chi^2$ values for $\Delta\xi_H$, $\Delta\xi_C$,
$\xi_{NH}$, $\xi_{NC}$, $\xi_{SH}$ and $\xi_{SC}$ in the $Q-W$ map
change with number of simulated maps used (the $Q-V$ and $V-W$ maps
produced very similar results). The results show clear convergence to
some value well within the $1\sigma$ confidence levels. The reason
why we observe faster convergence in these maps could simply be due to
the fact that we are removing the cosmological signal from the
analysis and with it much of the variance.

\begin{figure}
\vspace{0.6cm}
\begin{center}
\includegraphics[width=2.3in]{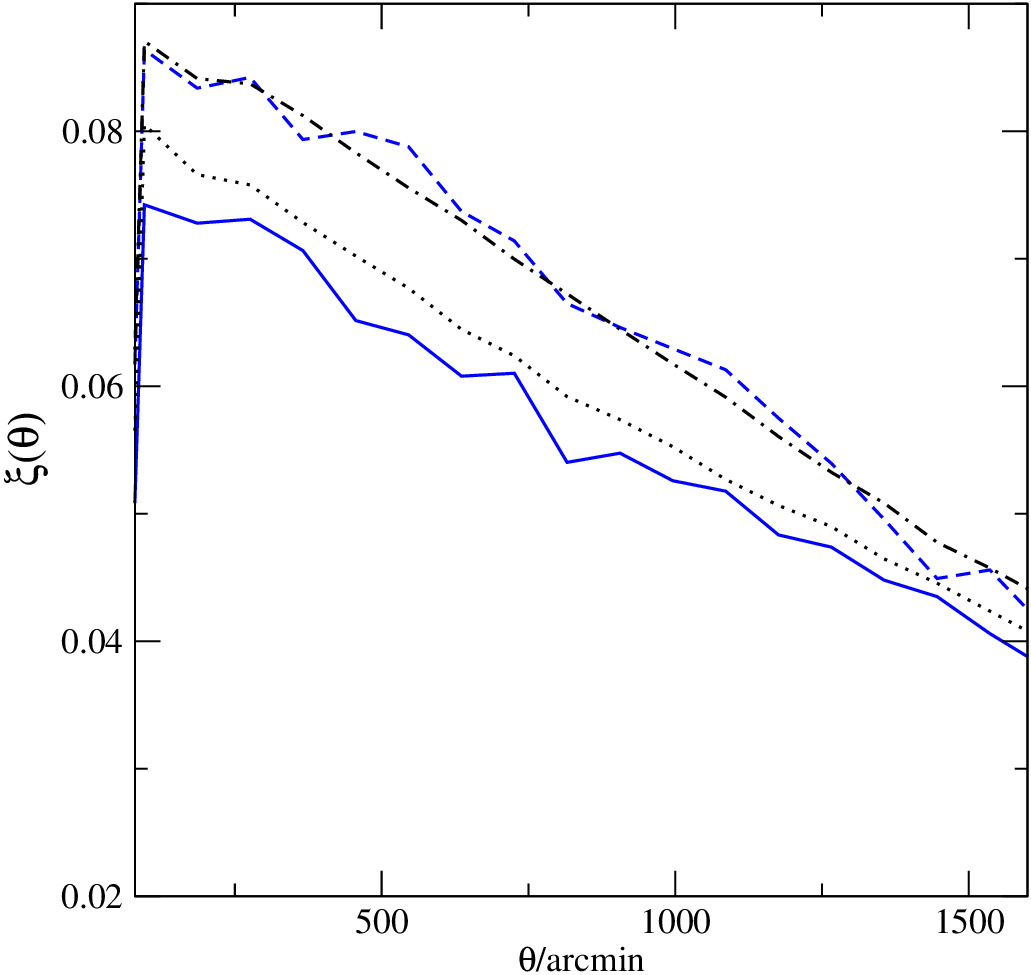}
\caption{\small $\xi_{NC}$ and $\xi_{SC}$ estimated from the
$V-Q$ subtracted maps. WMAP's data are the solid line for Southern
hemisphere and dashed line for Northern hemisphere (both in blue). Gaussian averaged
data in dotted line for Southern hemisphere, dot dashed line
for Northern hemisphere (both in black).
}
\label{fig:VQ_corr_NS}
\end{center}
\end{figure}

Figure \ref{fig:VQ_corr_NS} shows $\xi_{NC}$ and $\xi_{SC}$ for the
WMAP data and also $\bar{\xi}^G_{NC}$ and $\bar{\xi}^G_{SC}$ where the
average is done over the 110 simulated $V-Q$ noise maps.

Some comments on this figure are appropriate.  Firstly we note that there is an intrinsic
North/South asymmetry in the Gaussian noise maps. This is due to the
large-scale structure generated in the noise due to the uneven
scanning pattern of the WMAP satellite. We recall that pixel noise is
weighted according to the number of times a pixel has been observed, and as
such this feature is fully simulated in all our previous
maps. This
large-scale structure combined with the fact we are applying an
asymmetric mask to the data results in the non-zero and North/South
asymmetric peak-peak correlation function we see. We draw attention to the fact
that this asymmetry is qualitatively different
from what we found in Sections 4.2, 4.3, 4.4 and 4.5, since we now
find an excess in correlation in the Northern hemisphere, as opposed
to in the South\footnote{As a sanity test, we have also performed an identical
analysis on purely white noise maps which include the WMAP's satellite
scanning pattern and found them to have the same
North/South asymmetry behaviour.}. This excess in correlation in the North is indeed seen in the
Gaussian-averaged peak-peak correlation function of all our previous
maps, although on a much smaller scale. Finally we note that there is
a more noticeable deviation of the WMAP data from the
Gaussian simulations in the Southern hemisphere. However, we find none
of these to be significant. In fact, this statement extends to the
other two cases: $V-W$ and $Q-W$. We find no signs of non-Gaussianity
in any of the estimators in any of our combined noise and foreground
maps, with all the $\chi^2$ values well within values which are
consistent with the Gaussian hypothesis (our most extreme $\chi^2$
value comes from $\xi_{SH}$ in the $V-Q$ map, where we find
$\chi^2=1.49$ - see Table 2 in the next section for a summary of the
most extreme values in all three maps).

\subsection{A summary of our results}
\label{sec:summary}

In this subsection we take the opportunity to summarise our results
into one table and to elaborate on the confidence levels we have
quoted throughout the paper. We do this by presenting a table with all
the statistics for which we have found the WMAP data to have a reduced
$\chi_{WMAP}^2
\ge 2$, Table \ref{tab:summary}. 

We recall that in all cases we have rebinned the data into 19 linearly-spaced bins,
of which we use the last 18 to compute each of the $\chi^2$
statistics. The $P_{theory}$ column gives the probability of randomly
obtaining a given value of $\chi^2 \ge \chi^2_{WMAP}$ assuming the underlying distribution
is a $\chi^2$ distribution with 18 degrees of freedom,
and the $N_{Gaussian}$ column shows
how many Gaussian maps have a $\chi^2 \ge \chi^2_{WMAP}$ for the
correspondent estimator (the number in brackets in the total number of
Gaussian maps). It is worth noting that the $\chi^2$
distribution we estimate from the Gaussian maps fits a $\chi^2$
distribution with 18 degrees of freedom which has been shifted
slightly by $\Delta\chi^2\approx0.1$ to
lower values. Hence any limit on high values of $\chi^2$ based on this theoretical distribution
is a conservative one. These are the limits we quote throughout the
text. Shifting the Gaussian $\chi^2$ distribution by
$\Delta\chi^2=0.1$ results in the $P_{theory}$ values in Table 2
roughly being halved. 

We draw attention to our most striking detections, which come from the cold spots in the Southern
hemisphere, appearing both in the co-added QVW map and in the single frequency
Q band map with reduced $\chi^2$ values of $3.877$ and $3.831$
respectively. 

\begin{table*}
\begin{minipage}{110mm}
\caption{\small Our main detections. We present all situations which
yielded vales of $\chi^2 \ge 2$. In addition to this and for the sake of
completeness we also present the most extreme $\chi^2$ values obtained
in Section \ref{sec:differences}. $P_{theory}$ is the theoretical
probability of randomly obtaining a reduced $\chi^2 \ge \chi^2_{WMAP}$
assuming a reduced $\chi^2$ distribution with 18 degrees of freedom and
$N_{Gaussian}$ is the total number of Gaussian maps with $\chi^2 \ge \chi^2_{WMAP}$. In brackets is the number of Gaussian realisations used for each statistic.}
\begin{center}
\begin{tabular}{|c|c|c|c|c|c|} \hline
Map & $\ell_{cut}$  & Estimator         & $\chi^2_{WMAP}$   & $P_{theory}$ & $N_{Gaussian}$  \\ \hline
QVW & 10            & $\Delta\xi_C$     & 2.302      &  $1.32\times10^{-3}$  &  0 (250) \\ 
QVW & 5             & $\xi_{SC}$        & 2.358      &  $9.58\times10^{-4}$  &  0 (250) \\
QVW & 10            & $\xi_{SC}$        & 3.877      &  $4.91\times10^{-8}$  &  0 (250) \\
QVW & 20            & $\xi_{SC}$        & 2.747      &  $9.15\times10^{-5}$  &  0 (250) \\
QVW & 25            & $\xi_{SC}$        & 2.764      &  $8.23\times10^{-5}$  &  0 (250) \\            
QVW & 30            & $\xi_{SC}$        & 2.756      &  $8.65\times10^{-5}$  &  0 (250) \\ \hline
QVW & 10            & $\tilde{\xi}_C$   & 3.011      &  $1.71\times10^{-5}$  &  0 (250) \\
QVW & 20            & $\tilde{\xi}_C$   & 2.658      &  $1.59\times10^{-4}$  &  0 (250) \\
QVW & 25            & $\tilde{\xi}_C$   & 2.923      &  $3.01\times10^{-5}$  &  0 (250) \\
QVW & 30            & $\tilde{\xi}_C$   & 2.601      &  $2.25\times10^{-4}$  &  0 (220) \\ \hline
Q   & 10            & $\Delta\xi_C$     & 2.081      &  $4.57\times10^{-3}$  &  2 (250) \\ \hline
Q   & 10            & $\xi_{SC}$        & 3.831      &  $6.78\times10^{-8}$  &  0 (250) \\
V   & 10            & $\xi_{SC}$        & 2.571      &  $2.70\times10^{-4}$  &  0 (250) \\
W   & 10            & $\xi_{SC}$        & 2.729      &  $1.02\times10^{-4}$  &  0 (250) \\ \hline
Q   & 10            & $\tilde{\xi}_C$   & 2.156      &  $3.02\times10^{-3}$  &  0 (250) \\
V   & 10            & $\tilde{\xi}_C$   & 2.695      &  $1.26\times10^{-4}$  &  0 (250) \\
W   & 10            & $\tilde{\xi}_C$   & 2.325      &  $1.16\times10^{-3}$  &  0 (250) \\ \hline
Q   & 10            & $\tilde{\xi}_H$   & 2.029      &  $6.04\times10^{-3}$  &  0 (250) \\
W   & 10            & $\tilde{\xi}_H$   & 2.215      &  $2.17\times10^{-3}$  &  1 (250) \\ \hline
V-Q & 10            & $\xi_{SH}$        & 1.494      &  $8.10\times10^{-2}$  & 10 (110) \\
Q-W & 10            & $\Delta\xi_H$     & 1.328      &  $1.58\times10^{-1}$  & 22 (110) \\
V-W & 10            & $\Delta\xi_C$     & 1.426      &  $1.07\times10^{-2}$  & 10 (110) \\
\hline
\end{tabular}
\end{center}
\label{tab:summary}
\end{minipage}
\end{table*} 

%%%%%%%%%%%%% Section V %%%%%%%%%%%%%%%%%%%

\section{Conclusions}
\label{sec:conclusions}        

In this paper we have used the peak-peak correlation function of the
local extrema in the CMB temperature fluctuations to probe for
non-Gaussian signatures. As explained at the start of Section 4, we
have constructed a series of statistics which look both at cold and
hot spots in the full sky, in the Northern and Southern hemispheres
separately and at the difference between both hemispheres. We have
also looked at a variety of maps: a co-added QVW map with a standard
mask applied, a co-added QVW map with an extended galactic cut
applied, three co-added maps at the same frequency and three
differences maps so as to remove the cosmological signal. In the case
of the co-added QVW map we also investigated eight different cases
with some of the low multipoles removed. 

Our main results are summarised in Table \ref{tab:summary}
in Section \ref{sec:summary} - we find strong evidence for
non-Gaussianity, mainly associated with the cold spots and with the
Southern hemisphere; this non-Gaussianity disappears completely if we filter out
the harmonic modes $\ell \le 40$ and at least partially if we exclude
sky within $|b|<30^\circ$, so it is a large-scale effect associated
with the galactic plane.

Recently, \citet{art:LarsonAndWandelt05} have also
used the peak-peak correlation function of cold and hot spots in their
search for non-Gaussianity. Direct comparison of results is not
straightforward as the resolutions of the maps used in the two studies
are significantly different. However, in the simplest case where both
groups looked at the full sky CMB temperature field (with equivalent masks based on the
standard kp0 mask applied), both results are in agreement in the sense
that both fail to yield a detection. We believe this lack of detection
is a result of large cosmic variance in low-$\ell$ multipoles.

We investigate this further by removing some of the low
order multipoles from the maps, in the hope that by doing so we are
increasing our sensitivity to non-Gaussian features by reducing the
effects of cosmic variance. Once we remove all harmonic modes with $\ell \le 10 $ we
systematically find anomalies related to the cold spots in the WMAP
data and, when looking at both hemispheres separately, we not only
find a striking North/South asymmetry, we repeatedly find the
strongest anomalies to be in the Southern hemisphere. This is not
unheard of: \citet{art:VielvaEtAl03} first found an anomalous large
cold spot in the Southern hemisphere (nicknamed {\it The Spot}), detection which was followed by
\citet{art:CruzEtAl04}, \citet{art:MukherjeeWang04} and \citet{art:McEwenEtAl04} and confirmed repeatedly. However, we do find that
our detections disappear when we exclude sky regions within 30 degrees
of the galactic place (we recall that {\it The Spot} is localised at
approximately $(b=-57^\circ,l=209^\circ)$, well outside our cut
regions of sky). We therefore conclude that our detections come mainly
from something other than {\it The Spot}. 

The North/South asymmetry is also something which has been quoted time
and time again in the literature: \citet{art:Park03}, \citet{art:EriksenEtAl03,art:EriksenEtAl04,art:EriksenEtAl04b}
and other work previously quoted in this paper have consistently found
non-Gaussian asymmetries between the two galactic hemispheres. The
asymmetry we find in this study seems to be a large scale effect, once again related only
to the cold spots and to be contained within 30 degrees of the
galactic plane. 

We investigate our detections further by firstly conducting an analysis
in single frequency maps. We find some evidence for a dependence of
the signal with frequency when we look at different hemispheres
(peaking at 41GHz, corresponding to the Q band and in agreement with
\citet{art:LiuZhang05}), but this detection does not appear in a full-sky analysis. Secondly we
remove the cosmological signal from the analysis by subtracting
different frequency maps and testing the resulting foreground/noise
combination maps for non-Gaussian signals. We find no signs of
non-Gaussianity in these subtracted maps.

How do we make sense of these results?  A simple explanation seems
untenable.  The fact that the signal becomes insignificant when the
galactic plane is removed suggests unsubtracted galactic foregrounds are
responsible; the large-scale nature of the signal is certainly consistent
with this picture.  One would then expect the individual frequency maps to
show a significant signal, and this we do find, most strikingly in the Q
band.  However, the difference maps do not show a significant detection;
these maps should directly test the residual foregrounds and noise, so the
absence of detected non-Gaussianity does not obviously support this
picture.  We can reconcile these observations if the residual foregrounds
affect more than one frequency band, and the subtraction removes the
contamination to some extent.  The fact that we find non-Gaussianity in
all the single-frequency bands adds some support to this complex picture.
In our view this is the most likely explanation for the results we find,
but we cannot exclude a primordial origin for at least part of the
non-Gaussian signal.

%%%%%%%%%%%% Acknowledgments %%%%%%%%%%%%%%%%%%%

\section{Acknowledgments}
RT is funded by the Funda\c{c}\~{a}o para a Ci\^{e}ncia e a
Tecnologia under the reference PRAXIS SFRH/BD/16973/04. PGC is
supported by the PPARC through a Postoctoral Rolling grant. SG is
supported by the PPARC. SG and RT would like to thank 
Rob Crittenden for helpful discussions. RT would like to thank John
Peacock for helpful discussions and suggestions.
We acknowledge the use of the Legacy Archive for Microwave Background
data analysis (LAMBDA). Support for LAMBDA is provided by the NASA Office of Space Science. Some of the results in this paper have been
derived using the HEALPix (G\'{o}rski, Hivon, and Wandelt 1999) package.

%%%%%%%%%%%%% Biblio %%%%%%%%%%%%%%%%%%%

\bibliographystyle{mn2e}

\end{document}